\documentclass[twocolumn,aps,pra,showpacs,superscriptaddress,floatfix]{revtex4}

\usepackage{bm}
\usepackage{dsfont}
\usepackage{amsmath}
\usepackage{graphicx}
\usepackage{dcolumn}
\newcommand{\addrDresden}{Institut f\"{u}r Theoretische Physik, 
Technische Universit\"{a}t Dresden, 01062 Dresden, Germany}
\newcommand{\addrParis}{Laboratoire Kastler Brossel, \'Ecole Normale Sup\'erieure et
Universit\'{e} Pierre et Marie Curie, Case 74,
75005 Paris, France}
\newcommand{\addrGaithersburg}{National Institute of Standards and Technology, 
Mail Stop 8401, Gaithersburg, MD 20899-8401, USA}

\newcommand{\refPar}[1]{(\ref{#1})}
\newenvironment{newA}{}{}
\newcommand{\defi}{=} %
\newcommand{\complexI}{i} %
\newcommand{\landauO}{{\mathcal O}}
\newcommand{\DeltaFS}{\Delta_{\text{fs}}}
\newcommand{\funcExple}{J} %
\newcommand{\approxA}{\mathcal{A}_3} %
\newcommand{\coefApprox}[1]{a_{#1}} %
\newcommand{\coefApproxLJ}[1]{\coefApprox{#1}(l_j)} %

\begin{document}

\title{Perturbation Approach to the Self Energy of non-S Hydrogenic States}

\author{Eric-Olivier Le~Bigot}
\affiliation{\addrParis}
\affiliation{\addrGaithersburg}

\author{Ulrich D. Jentschura}
\affiliation{\addrParis}
\affiliation{\addrDresden}

\author{Peter J. Mohr}
\affiliation{\addrGaithersburg}

\author{Paul Indelicato}
\affiliation{\addrParis}

\author{Gerhard Soff}
\affiliation{\addrDresden}

\begin{abstract}
  We present results on the self-energy correction to the energy
  levels of hydrogen and hydrogenlike ions. The self energy represents
  the largest QED correction to the relativistic (Dirac-Coulomb) energy
  of a bound electron.  We focus on the perturbation expansion of the self energy
  of non-S states, and provide estimates of
  the so-called $A_{60}$ perturbative coefficient, which can be
  considered as a relativistic Bethe logarithm.  Precise values of $A_{60}$ are given for many P, D, F and G states, while estimates are given for other electronic states.
 These results can be used in high-precision
  spectroscopy experiments in hydrogen and hydrogenlike ions.  They
  yield the best available estimate of the self-energy correction of
  many atomic states.
\end{abstract}


\maketitle

\typeout{Section: Introduction}
\section{Introduction}
\label{SecIntro}

\begin{newA}

The recent dramatic progress in
high-precision spectroscopy (see, e.g.,~\cite{biraben2001}) has
motivated the calculation of numerous contributions to the energy
levels of hydrogen and hydrogenlike systems. Such spectroscopic
experiments test our understanding of atomic levels, and provide
precise determinations of fundamental constants~\cite{mohr2000b}; this
requires accurate predictions of atomic energies, and, in particular,
the calculation of corrections due to Quantum Electrodynamics (QED),
the quantum field theory of electromagnetic interactions.  The largest correction to the relativistic
(Dirac) energy levels of hydrogen
and hydrogenlike ions is provided by the so-called \emph{self-energy}
contribution of QED. The self energy is a process which modifies the
relativistic (Dirac) energy of an electron, and which can be depicted
by the following Feynman diagram,
\[
\makebox[0cm][c]{
\includegraphics{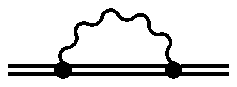},
}
\]
where the double line denotes the electron (bound to the nucleus), and
where the wavy line represents the photon emitted and reabsorbed by
the electron.  The self-energy correction to energy levels in hydrogen
and hydrogenlike ions can be expressed as an expansion in $Z\alpha$
and $\ln(Z\alpha)$ (see, e.g., \cite{ErYe1965a})---$Z$ is the nuclear
charge number of the nucleus of the hydrogenlike ion under
consideration, and $\alpha$~is the fine-structure constant.  Analytic
calculations of the (one-loop) self energy in bound systems have a
long history, starting from Bethe's seminal paper~\cite{Be1947}, and
have since extended over more than five decades.

The purpose of this paper is to provide good approximate values of the
self-energy correction to the energy levels of hydrogen and
hydrogenlike ions, for any P~state, and any state with a higher
angular momentum.  Only a part of the perturbation expansion of the
self energy of these states is known analytically.  The first two
non-analytically-known contributions to this expansion are the Bethe
logarithm $\ln k_0(nl) $ and the so-called~$A_{60}(nl_j)$ coefficient
of the self energy, which can be characterized as a
\emph{relativistic} Bethe logarithm [see Sec.~\ref{notations}, and in
particular Eqs.~\refPar{ESEasF}, \refPar{defFLOL1}
and~\refPar{A40gen}]. Here, $nl_j$ is the standard spectroscopic
notation for an atomic state.  This paper thus contains formulas for
estimating both of these important quantities (see
Sec.~\ref{sec:a60higherN} and~\ref{sec:approxA60andBL}).

Very precise numerical values of the Bethe logarithm $\ln k_0(nl) $
have been obtained (see, e.g., Refs.~\cite{goldman2000} and~\cite{drake90note}), and numerical
convergence acceleration techniques~\cite{AkSaJeBeSoMo2003} can  yield very precise values of this quantity for any atomic
state~$nl$.  The estimate~\refPar{eq:blCoefs}
that we obtained as a by-product in Sec.~\ref{sec:approxA60andBL} should be useful to
experiments that use levels for which no published values
of the Bethe logarithm exist~(see, e.g., Ref.~\cite{devries2002}).

Many new values of the relativistic Bethe logarithm~$A_{60}(nl_j)$
have recently been published~\cite{jentschura2003b}.  Other values
have been obtained previously for some
S~\cite{pachucki92,Pa1993,JeMoSo1999} and
P~states~\cite{jentschura96,JeSoMo1997}. This paper contains two
additional values [$A_{60}(5F_{5/2})$ and $A_{60}(5F_{7/2})$], as well
as details on the procedure that we used in obtaining the values
of~$A_{60}$ in Ref.~\cite{jentschura2003b} and in Table~\ref{tbl:a60F}
(see Sec.~\ref{ResHighLow}).

The results of Sec.~\ref{ResHighLow}--\ref{sec:approxA60andBL} provide
an improvement over the available approximations of the bound-electron
self energy, over a large range of nuclear charge numbers~$Z$. In
particular, they yield the best available estimates for the
self-energy correction in hydrogen, for all the states for which no
exact (non-perturbative) value of the self energy has yet been
published (i.e., all levels, except $n=1$ and $n=2$
levels~\cite{JeMoSo1999,jentschura2001b}).

It is important to know accurately the energy (and in particular the
self energy) of higher angular momentum states, because they are used
in high-precision spectroscopic
measurements~\cite{NiEtAl2000,debeauvoir2000,ReEtAl2000,schwob99,schwob99b,BeEtAl1997}.
States with very-high angular orbital quantum numbers~$l\simeq 30$
have been recently used in such experiments~\cite{devries2002}.
Further motivation for the present study results from the need to
accurately compare the two approaches that have been used for the
theoretical study of QED shifts, so as to check their consistency: (i)~the analytic expansion in the parameter $Z\alpha$%
, mostly used for low-$Z$ systems, and (ii)~the numerical approach,
which avoids the $Z\alpha$ expansion and has been used predominantly
for the theoretical description of high-$Z$ hydrogenlike ions~\cite{mohr98}.

Recently, the most accurate methods implementing a non-perturbative
calculation of the self
energy~\cite{jentschura2001b,jentschura99d,indelicato98,mohr74,mohr74b} have
been extended by analytic results~\cite{LBInMo2001}. Taken
together, they provide access to the self-energy shift of electrons of
total angular momentum $j >
3/2$. %
This has allowed us to obtain numerical values of the self energy, and
to use them in \emph{checks} of the~$A_{60}$ coefficients presented in
Tables~\ref{tbl:a60P}--\ref{tbl:a60G} (see Sec.~\ref{sec:check}).

Moreover, general progress in theoretical calculations of atomic
energy levels has been achieved by means of numerical
algorithms~\cite{jentschura99d,JeMoSoWe1999,AkSaJeBeSoMo2003} that lead to an
accelerated convergence of the angular-momentum series expansion of
the bound-electron relativistic Green function.  Such algorithms are
also useful for performing the series summations that we had to do in
order to obtain the values of~$A_{60}$ presented here (see
Sec.~\ref{ResHighLow}).

Notation and conventions are defined in Sec.~\ref{notations}. The
mathematical method used for the semi-analytic calculations of
$A_{60}$ in Ref.~\cite{jentschura2003b} is discussed in
Sec.~\ref{EpsilonMethod}.  Details on these calculations are presented
in Sec.~\ref{ResHighLow}.  Formulas for the relativistic Bethe
logarithm~$A_{60}(nl_j)$ of P and D~states are presented in
Sec.~\ref{sec:a60higherN}\@. Estimates of the Bethe logarithm $\ln
k_0(nl) $ and of~$A_{60}(nl_j)$ as a function of the orbital quantum
number~$l$ are reported in Sec.~\ref{sec:approxA60andBL}.  We have
performed additional checks of the values of~$A_{60}$ in
Tables~\ref{tbl:a60P}--\ref{tbl:a60G}, as described in
Sec.~\ref{sec:check}; we also show in that section that for the states considered here, the inclusion
of~$A_{60}$ in the (truncated) perturbation expansion of the electron
self energy [Eq.~\refPar{defFLOL1} below] does indeed improve the self energy estimates. A summary
of the paper is given in Sec.~\ref{SummaryOfResults}. The fitting
method that we used in obtaining asymptotic behaviors of~$\ln k_0(nl) $
and of~$A_{60}(nl_j)$ is described in the Appendix.

\end{newA}

\typeout{Section: Notation and Conventions}
\section{Notation and Conventions}\label{notations}

In this section,
we define the notation and conventions used in this paper.
We write the (real part of the) one-loop self-energy shift 
of an electron in the level~$n$ with orbital angular momentum~$l$ and total angular momentum~$j$
as
\begin{equation}
\label{ESEasF}
\Delta E_{\rm SE} = \frac{\alpha}{\pi} \, \frac{(Z \alpha)^4}{n^3} \, 
F(nl_j,Z\alpha) \, m \, c^2,
\end{equation}
where $F(nl_j,Z\alpha)$ is a dimensionless quantity. 
We use natural
units,
in which $\hbar = c = m = 1$ and $e^2 = 4\pi\alpha$
($m$ is the electron mass).
It is customary in the literature
to suppress
the dependence of $F$ on the quantum numbers $n$, $j$ and $l$
and write $F(Z\alpha)$ for $F(nl_j,Z\alpha)$.

The quantum numbers $l$ and~$j$ can be combined into the Dirac 
angular quantum number $\kappa$. As a function of $j$ and $l$,
$\kappa$ is given by
\begin{subequations}\label{eqs:defKappa}
\begin{equation}
\kappa = 2 \, (l-j) \, (j + 1/2),
\end{equation}
i.e.,
\begin{equation}
\kappa = - (j + 1/2) \quad \mbox{for} \quad j = l+1/2,
\end{equation}
and
\begin{equation}
\kappa = (j + 1/2) \quad \mbox{for} \quad j = l-1/2.
\end{equation}
\end{subequations}

The quantum numbers $j$ and $l$ 
can be derived from $\kappa$ according to
\begin{equation}
l = |\kappa + 1/2| - 1/2
\end{equation}
and 
\begin{equation}
j = |\kappa| - 1/2,
\end{equation}
i.e., $\kappa$ specifies uniquely both $j$ and $l$.
The semi-analytic expansion of $F(nl_j,Z\alpha)$
about $Z\alpha = 0$ for a general atomic state with quantum numbers $n$,
$l$ and $j$ gives rise to the expression~\cite{ErYe1965a}
\begin{eqnarray}
\label{defFLO}
F(nl_j,Z\alpha) &=& 
A_{41}(nl_j) \, \ln[(Z \alpha)^{-2}] \nonumber\\[2ex]
& & + A_{40}(nl_j) +
(Z \alpha) \, A_{50}(nl_j) \nonumber\\[2ex]
& & + \, (Z \alpha)^2 \,
\bigl\{A_{62}(nl_j) \, \ln^2[(Z \alpha)^{-2}] 
\nonumber\\[2ex]
& & + A_{61}(nl_j) \,\ln[(Z \alpha)^{-2}] 
\nonumber\\[2ex]
& &  + G_{\rm SE}(nl_j,Z\alpha) \bigr\}.
\end{eqnarray}
This expansion is semi-analytic, i.e., it involves powers of
$Z\alpha$ and of~$\ln[(Z\alpha)^{-2}]$.  Terms added to the leading
order in $Z\alpha$ are commonly referred to as the binding
corrections.
The $A$ coefficients have two indices, the first of which denotes the
power of $Z\alpha$ [including those powers contained in
Eq.~(\ref{ESEasF})], while the second index denotes the power of the
logarithm $\ln(Z \alpha)^{-2}$.

The limit as $Z\alpha \to 0$ of $G_{\rm SE}(nl_j,Z\alpha)$ is known to
be finite and is referred to as the $A_{60}$ coefficient, i.e.,
\begin{equation}\label{Asixty}
A_{60}(nl_j) \defi \lim_{Z\alpha \to 0} G_{\rm SE}(nl_j, Z\alpha).
\end{equation}
Historically,
the evaluation of the coefficient $A_{60}$ has been highly problematic.
Due to the large number of terms that contribute at relative order 
$(Z\alpha)^2$ in~\refPar{defFLO} and problems concerning the 
separation of terms that contribute to a specific order
in the $Z\alpha$ expansion, evaluations are plagued with severe  
calculational as well as conceptual difficulties.
For example, the evaluation of
$A_{60}(1{\rm S}_{1/2})$ has drawn
a lot of attention for a long time~\cite{ErYe1965a,ErYe1965b,%
Er1971,Sa1981,Pa1993}. In general, the complexity of the calculation
increases with increasing principal quantum number~$n$.

For many states, some of the coefficients in (\ref{defFLO})
vanish. Notably,
this is the case for P states and for states with higher angular
momenta, as a consequence of their behavior at the nucleus, which is
less singular than that of S~states (specifically, we have $A_{62}(nl_j) =
A_{50}(nl_j) = A_{41}(nl_j) = 0$ for
$l \neq 0$---see Refs.~\cite{ErYe1965a,ErYe1965b} and references
therein).  The fact that the logarithmic coefficient $A_{71}(nl_j)$ contained in $G_{\text{SE}}(nl_j,Z\alpha)$ in~\refPar{defFLO}
vanishes for $l \neq 0$ has been pointed out in~\cite{Ka1997}; it is therefore expected that~$A_{7k}(nlj)=0$ for~$k>1$.  For
nonzero~$l$, we thus have
\begin{eqnarray}
\label{defFLOL1}
F(nl_j,Z\alpha) & = & A_{40}(nl_j) + (Z \alpha)^2 \,
\left[A_{61}(nl_j) \, \ln(Z \alpha)^{-2} \right. 
\nonumber\\[2ex]
& & \left. + A_{60}(nl_j) \right] + 
\landauO[(Z\alpha)^3] \qquad (l \neq 0).
\end{eqnarray}
For the comparison to experimental data,
it is useful to note that the terms in
(\ref{defFLO}) and (\ref{defFLOL1}) acquire 
reduced-mass corrections according to Eqs.~(2.5a) 
and (2.5b) of~\cite{SaYe1990}.

The general formula for $A_{40}$ for a non-$\text{S}$ state reads (see, e.g., \cite{mohr2000b,SaYe1990,ErYe1965a})
\begin{equation}
\label{A40gen}
A_{40}(nl_j) = -\frac{1}{2\kappa\,(2 l + 1)} - \frac{4}{3}\,\ln k_0(nl) ,
\end{equation}
where the Bethe logarithm $\ln k_0(nl) $ is an inherently
nonrelativistic quantity, whose expression reads~\cite[\S~19]{bethe57withNote}
\begin{eqnarray}
\label{bethelog}
\lefteqn{\ln k_0(nl)  = \frac{n^3}{2 (Z\alpha)^4\,m} }
\\\nonumber
& & \quad \times \left< \phi \left| \frac{p^i}{m} \, 
\left( H_{\rm S} - E_n \right) \,
\ln \left[ 2\frac{\left| H_{\rm S} - E_n \right|}{(Z\alpha)^2\,m} \right] \,
\frac{p^i}{m} \right| \phi \right> . 
\end{eqnarray}
Here, $H_{\rm S}$ is the nonrelativistic Coulomb Hamiltonian $\bm{p}^2/(2m) -
(Z\alpha)/r$, $p^i$ are the components of the momentum operator ($i$
is summed over from 1 to~3), and the ket $|\phi\rangle$ represents the
Schr\"{o}dinger wavefunction of a state with quantum numbers $(n,l)$,
with associated bound-state energy $E_n = -(Z\alpha)^2 \, m/(2 \,
n^2)$.  
The Bethe logarithm
is spin-independent and therefore independent
of the total angular momentum $j$ for a given orbital angular momentum
$l$; it can be written as a function of $n$ and $l$ alone [factors of $Z$ cancel out in Eq.~\refPar{bethelog}, so that the Bethe logarithm does not depend on~$Z$]. 
For the atomic levels under investigation here,
the Bethe logarithm has been evaluated 
in Refs.~\cite{BeBrSt1950,Ha1956,ScTi1959,Li1968,Hu1969,KlMa1973,haywood85,drake90note,%
FoHi1993,goldman2000} (the results exhibit varying accuracies).
Because $A_{60}$ involves relativistic corrections to the 
coefficient $A_{40}$, which in turn contains
the Bethe logarithm, it is natural to refer
to $A_{60}$ as a ``relativistic Bethe logarithm.''

A general analytic result for the logarithmic correction
$A_{61}$ as a function of the bound state quantum numbers $n$, $l$ and
$j$ can be inferred from Eq.~(4.4a) of~\cite{ErYe1965a,ErYe1965b} upon
subtraction of the vacuum-polarization contribution 
contained in the quoted equation. We have
\begin{eqnarray}
\label{A61gen}
A_{61}(nl_j) &=& \frac{4}{3} \Biggl\{
\frac{8 \, \left(1 - \delta_{l,0}\right) \,
\left(3 - \frac{\displaystyle l(l+1)}{\displaystyle n^2} \right)}
  {\prod\limits_{m=-1}^3 (2\,l + m)} \\
&& + \delta_{l,1} \,
\biggl(1-\frac{1}{n^2}\biggr) \, \biggl(\frac{1}{10} + \frac{1}{4} \,
\delta_{j,l-1/2} \biggr) \nonumber\\
&& + \delta_{l,0} \, \biggl[-\frac{601}{240} - \frac{77}{60 \, n^2} 
\nonumber\\
& & + 7 \ln 2 + 3 \left( \gamma - \ln n + \Psi(n+1) \right) \biggr] 
\Biggr\}.
\nonumber
\end{eqnarray}
Here, $\Psi$~denotes the logarithmic derivative of the $\Gamma$
function~\cite[\S~6.3]{abramovitz72}, and $\gamma$~is Euler's constant~\cite[\S~6.1.3]{abramovitz72}.
We may infer immediately
\begin{subequations}\label{eq:A61partic}
\begin{eqnarray}
\label{A61P12}
A_{61}(nP_{1/2}) &=& \frac{1}{45}\, \left(33 - \frac{29}{n^2}\right),
\\[2ex]
\label{A61P32}
A_{61}(nP_{3/2}) &=& \frac{2}{45}\, \left(9 - \frac{7}{n^2}\right),
\\[2ex]
\label{A61L2}
A_{61}(nl_j) &=& 
\frac{32 \, \left(3 - \frac{\displaystyle l\,(l+1)}{\displaystyle n^2} \right)}
{3 \prod\limits_{m=-1}^3 (2\,l + m)} \quad 
(l \geq 2).  
\end{eqnarray}
\end{subequations}
For a given orbital angular momentum $l$, the coefficient $A_{61}$ 
approaches a constant as $n \to \infty$.
Equation (\ref{A61L2}) implies that $A_{61}$ is spin-independent for $l \geq 2$,
i.e., for D, F, G, $\dots$ states. Therefore, $A_{61}$ does not contribute
to the fine structure of these states.  

\typeout{Section: The $\epsilon$ Method}

\section{The $\epsilon$ Method}
\label{EpsilonMethod}

In this section, we illustrate the usefulness of the so-called
$\epsilon$ method~\cite{Pa1993,jentschura96,JeSoMo1997} in bound-state
calculations of QED corrections.  It is known that
relativistic corrections to the wavefunction and  higher-order terms in the expansion of
the bound-electron propagator in powers of Coulomb vertices
 generate QED corrections of higher
order in $Z\alpha$ (see, e.g., Ref.~\cite{pachucki93} and references
therein); these terms manifest themselves in Eq.~(\ref{defFLO}) in the
form of the function $G_{\rm SE}(nl_j,Z\alpha)$, which summarizes
these effects at the order of $\alpha\,(Z\alpha)^6\,m$---see Eqs.\ \refPar{ESEasF} and~\refPar{defFLO}.  It is also
well known that for very soft virtual photons, the potential expansion
fails and generates an infrared divergence, which is cut off by the
atomic momentum scale, $Z\alpha$ (see, e.g., Ref.~\cite{pachucki93}
and references therein). This cut-off for the {\em infrared}
divergence is one of the mechanisms that lead to the logarithmic terms
in Eq.~(\ref{defFLO}).

The $\epsilon$ method is used for the separation of the two different
energy scales for virtual photons: the nonrelativistic domain, in
which the virtual photon assumes values of the order of the atomic
binding energy, and the relativistic domain, in which the virtual
photon assumes values of the order of the electron rest mass.  We
consider here a model problem with one ``virtual photon,'' that
involves the separation of the function being integrated into a high- and 
a low-energy contribution.  This requires the temporary introduction
of a parameter $\epsilon$; the dependence on $\epsilon$ will cancel at
the end of calculation [see Eq.~(\ref{IHLresult}) below] when the
high- and the low-energy parts are added together. We have,
\begin{eqnarray}\nonumber
\mbox{nonrelativistic domain} & \ll \epsilon \ll & 
\mbox{electron rest mass,} \\\label{energyScales}
\text{i.e., } (Z \alpha)^2 \, m & \ll \epsilon \ll & m.
\end{eqnarray}
The high-energy part is associated with photon energies $\omega >
\epsilon$, and the low-energy part is associated with photon energies
$\omega < \epsilon$.

In order to illustrate the principles behind the $\epsilon$~method, we discuss a
simple, one-dimensional example: the evaluation of
\begin{equation}\label{defJ}
\funcExple(Z\alpha) \defi
\int_0^1 
\frac{(Z\alpha)^2 - \omega}{(Z\alpha)^2 + \omega} \,
\frac{1}{\sqrt{1 - \omega^2}} \, 
{\mathrm{d}}\omega,
\end{equation}
where the integration variable $\omega$ may be interpreted as the
``energy'' of a ``virtual photon.'' The integral $\funcExple$ can be explicitly
calculated, so that the perturbation expansion can be checked:
\begin{equation}\label{eq:exactJ}
\funcExple(Z\alpha) = -\frac{\pi}{2} + 
\frac{2 \, (Z\alpha)^2 \, 
\ln\left[ 
  {\displaystyle \frac{1}{(Z\alpha)^2}}
  \left(
    \sqrt{1-(Z\alpha)^4} + 1
  \right)
\right]
}
{\sqrt{1 - (Z\alpha)^4}}.
\end{equation}
For $|Z\alpha| < 1$, this formula is uniquely defined; for other values of~$Z\alpha$, the analytic
continuations of the logarithm and of the square-root have to be
performed consistently with the original definition (\ref{defJ}).

Within the $\epsilon$ method, we start by dividing the calculation of
$J(Z\alpha)$ into a high-energy part
$J_{\mathrm{H}}(Z\alpha,\epsilon)$ and a low-energy part
$J_{\mathrm{L}}(Z\alpha,\epsilon)$, each of which depends on an
additional parameter $\epsilon$ [that satisfies~(\ref{energyScales})].
The sum of the high- and low-energy contributions, which is
\begin{equation}\label{eq:jSum}
J(Z\alpha) = J_{\mathrm{H}}(Z\alpha,\epsilon) +
J_{\mathrm{L}}(Z\alpha,\epsilon),
\end{equation}
does not depend on~$\epsilon$. Thus, the dependence on~$\epsilon$
should vanish entirely when we add the high- and low-energy
contributions. We may therefore expand both contributions $J_{\rm H}$
and $J_{\rm L}$ first in $Z\alpha$, then in $\epsilon$, and then add
them up at the end of the calculation in order to obtain the
semi-analytic expansion of $J(Z\alpha)$ in powers of $Z\alpha$ and
$\ln(Z\alpha)$.

Let us first discuss the ``high-energy part'' of the calculation. 
It is given by the expression
\begin{eqnarray}
\label{IH}
\funcExple_{\mathrm{H}}(Z\alpha,\epsilon) &=& 
\int_\epsilon^1 
\frac{(Z\alpha)^2 - \omega}{(Z\alpha)^2 + \omega} \,
\frac{1}{\sqrt{1 - \omega^2}} \, 
{\mathrm{d}}\omega,
\end{eqnarray}
where it is important to note in particular the lower integration
limit,~$\epsilon$.
For $\omega > \epsilon$, we may expand
\begin{equation}
\label{expansion1}
\frac{ (Z\alpha)^2 - \omega }{ (Z\alpha)^2 + \omega} = 
-1 + \frac{2\,(Z\alpha)^2}{\omega} +
\landauO[(Z\alpha)^4]
\end{equation}
[see Eq.~\refPar{energyScales} with~$m=1$].
Each corresponding term of~\refPar{IH} can be integrated, with result
\begin{eqnarray}
\label{IHresult}
\lefteqn{\funcExple_{\mathrm{H}}(Z\alpha,\epsilon)=}&&
\\
\nonumber
&&\quad
\left(- \frac{\pi}{2}  + \ldots\right)
+ 2\, (Z\alpha)^2 \, \left[ \ln\left(\frac{2}{\epsilon}\right) + \ldots \right]
+ \landauO[(Z\alpha)^4],
\end{eqnarray}
where the ellipsis represent terms that vanish as~$\epsilon
\rightarrow 0$.
It is sufficient to only include terms that don't vanish as
$\epsilon\rightarrow 0$, to each order in $Z\alpha$, because the sum
$\funcExple$ in Eq.~\refPar{eq:jSum} does not depend on~$\epsilon$.
Moreover, this makes the calculation more manageable.
The full cancellation of the dependence on $\ln \epsilon$ will be
explicit after we evaluate the ``low-energy part.''

The contribution of the low-energy part ($0 < \omega < \epsilon$) reads
\begin{eqnarray}
\label{IL}
\funcExple_{\mathrm{L}}(Z\alpha,\epsilon) &=& 
\int_0^\epsilon 
\frac{(Z\alpha)^2 - \omega}{(Z\alpha)^2 + \omega} \,
\frac{1}{\sqrt{1 - \omega^2}} \, 
{\mathrm{d}}\omega,
\end{eqnarray}
where the upper limit of integration depends on~$\epsilon$.
For $\omega < \epsilon$, we use an expansion that avoids the infrared
divergences that we encountered in Eq.~(\ref{expansion1}):
\begin{equation}
\label{expansion2}
\frac{1}{\sqrt{1 - \omega^2}} = 1 + \frac{\omega^2}{2} + 
\frac{3}{8}\,\omega^4 + \cdots, %
\end{equation}
which leads to a~$Z\alpha$ expansion
of the low-energy part. We obtain for $\funcExple_{\rm L}$:
\begin{eqnarray}
\label{ILresult}
\lefteqn{\funcExple_{\mathrm{L}}(Z\alpha,\epsilon) = }&& 
\\
\nonumber
&&  (\ldots)  {}+ 2 \, (Z\alpha)^2 \, 
\left[\ln\frac{\epsilon}{(Z\alpha)^2} + \ldots \right]
 + \landauO[(Z\alpha)^4 \ln^j(Z\alpha)]  ,
\end{eqnarray}
where the ellipsis again represents terms that vanish as~$\epsilon
\rightarrow 0$, and where $j$~is some integer.

When the high-energy part~(\ref{IHresult}) and the low-energy
part~(\ref{ILresult}) are added,
the logarithmic divergences in~$\epsilon$
cancel, as it should, and we have
\begin{eqnarray}
\label{IHLresult}
\funcExple(Z\alpha) &=& \funcExple_{\mathrm{H}}(Z\alpha, \epsilon) + 
\funcExple_{\mathrm{L}}(Z\alpha, \epsilon) 
\nonumber\\[2ex]
&=& -\frac{\pi}{2} + 2 \, (Z\alpha)^2 \,
\left( \ln[(Z\alpha)^{-2}] + \ln 2 \right)
\nonumber\\[2ex]
& & + \landauO[(Z\alpha)^4 \ln^j(Z\alpha)]
\end{eqnarray}
(for some~$j$), which is consistent with~\refPar{eq:exactJ}.
We note the analogy of the above expression with the leading-order
terms of the $Z\alpha$ expansion of the function $F(nl_j, Z\alpha)$
given in Eq.~(\ref{defFLOL1}) for $l \neq 0$ (terms associated to the
coefficients $A_{40}$, $A_{61}$, and $A_{60}$).  In an actual Lamb
shift calculation, the simplifications observed between terms
containing~$\epsilon$ are crucial~\cite{jentschura96,JeSoMo1997}.

In this model example, the epsilon method allowed us to
obtain~\refPar{IHLresult} with minimal effort.  For comparison, the
reader may consider App.~A of~\cite{JePa2002}, which illustrates the
cancellation of $\epsilon$ in higher orders of the
$Z\alpha$ expansion, using a different example.

\typeout{Section: Calculation of Self-Energy Coefficients} 
\section{Calculation of Self-Energy Coefficients}
\label{ResHighLow}

This section, along with the previous one, gives detail on the methods
we used in order to obtain the values of the $A_{60}$ coefficient in
Tables~\ref{tbl:a60P}--\ref{tbl:a60G} (see also
Ref.~\cite{jentschura2003b}).  The purpose of our calculations is to
provide data for the self-energy coefficients up to and including the
relative order $(Z\alpha)^2$ [see Eq.~(\ref{defFLOL1})]; for the
states of interest here (non-S states) this corresponds to the
coefficients $A_{40}$, $A_{61}$ and $A_{60}$.  Equation (\ref{A40gen})
is the well-known general formula for the coefficient $A_{40}$.  The
coefficient $A_{61}$ can be found in Eq.~(\ref{A61gen}), with special
cases treated in Eqs.~(\ref{A61P12})--(\ref{A61L2}). The remaining
nonlogarithmic term $A_{60}$ is by far the most difficult to evaluate,
and the first results for any with orbital angular momentum quantum
number $l \geq 2$ were recently obtained in
Ref.~\cite{jentschura2003b} by using the methods described in this
section.

As explained in detail
in~\cite{Pa1993,jentschura96,JeSoMo1997}, the calculation
of the one-loop self energy falls
naturally into a high- and a low-energy part
($F_H$ and $F_L$, respectively).
In Sec.~\ref{EpsilonMethod}, we  illustrated 
this procedure, and the introduction of
the scale-separation parameter
$\epsilon$ for the photon energy.
According to~\cite[Eqs.~(39)--(43)]{jentschura96}, the 
contributions to the low-energy
part can be separated naturally into the nonrelativistic dipole and
the nonrelativistic quadrupole parts, and into relativistic corrections
to the current, to the Hamiltonian, to the binding energy and
to the wavefunction of the bound state. We follow here
the approach outlined in Refs.~\cite{jentschura96,JeSoMo1997},
with some  modifications.

One main difference as compared to
the evaluation scheme described previously concerns
the nonrelativistic quadrupole (nq) part.
It is given by  a specific
matrix element~(see the definition
of $P_{\rm nq}$ in Ref.~\cite[Eq.~(39)]{jentschura96}), which has to
be evaluated on the atomic state and averaged over the 
angles of the photon wave vectors:
\begin{eqnarray}
\label{pnq}
\int \frac{{\rm d} \Omega_k}{4 \pi}\,
P_{\rm nq} &=& 
\int \frac{{\rm d} \Omega_k}{4 \pi}\,
 \frac{\delta^{{\rm T},ij}}{6 m} 
\nonumber\\[2ex]
& & \!\!\!\!\!\!\!\!\!\!\!\!\!\!\!\!\!\!\!\! \!\!\!\!\!
\times \, \left[ \left< \phi \left| p^i \,
e^{ {\complexI}\, \bm{k} \, \cdot \, \bm{r}} \,
\frac{1}{H_S - (E - \omega)} \, p^j \,
e^{ - {\complexI} \bm{k} \, \cdot \, \bm{r}} \right| \phi \right> \right.
\nonumber\\[2ex]
& & \!\!\!\!\!\!\!\!\!\!\!\!\!\!\!\!\!\!\!\! \!\!\!\!\!
\left.
\quad \mbox{}- 
\left< \phi \left| p^i \,
\frac{1}{H_S - (E - \omega)} \, p^j \right| \phi \right> \right]
\end{eqnarray}
where the transverse $\delta$ function is given by 
\[
\delta^{{\rm T},ij} = \delta^{ij} - \frac{k^i\,k^j}{\bm{k}^2}.
\]
The dipole interaction obtained by the replacement
\[
\exp({\complexI}\, \bm{k} \, \cdot \, \bm{r}) 
\to 1
\]
is subtracted; it leads to a lower-order contribution.
The next term in the Taylor expansion of the exponential reads
\begin{eqnarray}
\label{FnqCoor}
\int \frac{{\rm d} \Omega_k}{4 \pi}\,
 \frac{\delta^{{\rm T},ij}}{6 m} 
\\[2ex]\nonumber
& & \!\!\!\!\!\!\!\!\!\!\!\!\!\!\!\!\!\!\!\! \!\!\!\!\!
\times \, \left[\left< \phi \left| p^i \,
(\bm{k} \, \cdot \, \bm{r}) \,
\frac{1}{H_S - (E - \omega)} \, p^j \,
(\bm{k} \, \cdot \, \bm{r}) \right| \phi \right>\right.
\\[2ex]\nonumber
& & \!\!\!\!\!\!\!\!\!\!\!\!\!\!\!\!\!\!\!\! \!\!\!\!\!
\left.
\quad \mbox{}- 
\left< \phi \left| p^i \,
\frac{1}{H_S - (E - \omega)} \, p^j \,
(\bm{k} \, \cdot \, \bm{r})^2 \right| \phi \right> \right].
\end{eqnarray}
This representation makes an evaluation in coordinate space possible.
However, an evaluation of this expression leads to a rather involved
angular momentum algebra. Specifically, we employ a well-known angular
momentum decomposition of the coordinate-space
hydrogen Green function~\cite{wichmann61}
\begin{equation}
\label{SCGreensf}
G(\bm{r}_1, \bm{r}_2, E - \omega) = \sum_{l',m} \,
g_{l'}(r_1, r_2, \nu) \,
Y_{l',m} \left(\hat{\bm{r}}_1\right) \,
Y_{l',m}^{*} \left(\hat{\bm{r}}_2\right),
\end{equation}
with $E - \omega \defi - \alpha^2 \, m / (2 \nu^2)$, and~\cite{hostler70}
\begin{eqnarray}
\label{gl}
\lefteqn{
g_{l'}(r_1, r_2, \nu) 
=
 \frac{4 m}{a \nu}
\left( \frac{2 r_1}{a \nu} \right)^{l'} \,
\left( \frac{2 r_2}{a \nu} \right)^{l'} \,
e^{- (r_1 + r_2)/(a \nu) }
}
\qquad \qquad \, &
\nonumber\\[2ex]
& & \times \sum_{k=0}^{\infty}
\frac{L_k^{2 l' + 1}\left(\frac{2 r_1}{a \nu} \right) \,
L_k^{2 l' + 1}\left(\frac{2 r_2}{a \nu} \right)}
{(k+1)_{2 l' + 1} \, (l' + 1 + k - \nu)},
\end{eqnarray}
where $a = 1 / (Z \alpha m)$, $(k)_c$ is the Pochhammer symbol, and
$L$ denotes associated Laguerre polynomials~\cite{abramovitz72}.
For a reference state
$|\phi\rangle$ of orbital angular momentum~$l$, we obtain in (\ref{FnqCoor})
nonzero contributions from Green-function components
(\ref{SCGreensf}) with $l' = l-2, l-1, l, l+1, l+2$.  They can be
obtained by straightforward, but tedious application of angular
momentum algebra (see, e.g.,~\cite{Ed1957}).

As in previous calculations (see also~\cite[Eqs.~(18) and~(19)]{jentschura96}
and~\cite[Eqs.~(55)--(58)]{JeSoMo1997}),
we obtain  for the high-energy part of all atomic states
the general structure
\begin{eqnarray}
\label{HighGen}
F_H (nl_j, Z\alpha) &=& - \frac{1}{2\kappa\,(2 l + 1)} 
\\\nonumber
& & + (Z\alpha)^2 \,
  \left[{\cal K} - \frac{{\cal C}}{\epsilon} -
    A_{61} \, \ln(2\epsilon) + \landauO(\epsilon) \right]
\\ \nonumber
&& + \ldots
\end{eqnarray}
where ${\cal K}$ is a constant, and where the ellipsis denotes
higher-order terms [in $Z\alpha$ and $\ln(Z\alpha)$].
As observed in Sec.~\ref{EpsilonMethod}, we may suppress terms
that vanish in the limit $\epsilon \to 0$ [terms
of the form $\landauO(\epsilon)$ in the $(Z\alpha)^2$-term in
Eq.~(\ref{HighGen}) above]. 
These terms cancel when
the high- and low-energy parts are added.

Together with the constant term $-A_{61} \, \ln 2$,
the constant ${\cal K}$ contributes to $A_{60}$. 
${\cal C}$ is the coefficient
of the $1/\epsilon$ divergence; the term 
$-{\cal C}/\epsilon$ cancels when the high- and low-energy parts
are added. Both ${\cal K}$ and ${\cal C}$ are
state dependent and vary with $n,j,l$. 
As in~\cite[Eqs.~(56) and~(57)]{jentschura96}
and~\cite[Eqs.~(89)--(92)]{JeSoMo1997},
the low-energy part, for all states under investigation,
has the general structure
\begin{eqnarray}
\label{LowGen}
\lefteqn{F_L (nl_j, Z\alpha)  =  - \frac{4}{3} \ln k_0(nl) }
\quad\quad&&
\\  \nonumber
& & {}+ (Z\alpha)^2 \, \left[{\cal L} + \frac{{\cal C}}{\epsilon} +
A_{61} \, \ln\left(\frac{\epsilon}{(Z \alpha)^2}\right)
+ \landauO(\epsilon) \right]
\\ \nonumber
&& {} + \ldots
\end{eqnarray}
where $\ln k_0(nl) $ is the Bethe logarithm [see Eq.~(\ref{bethelog})],
and where the ellipsis denotes higher-order terms.  The
cancellation of the divergence in $\epsilon$ between (\ref{HighGen})
and (\ref{LowGen}) is obvious.  The constant ${\cal L}$, which is
state-dependent (a function of $n,j,l$), represents the low-energy
contribution to $A_{60}$ and can be interpreted as the relativistic
generalization of the Bethe logarithm.  In terms of the general
expressions (\ref{HighGen}) and (\ref{LowGen}), $A_{60}$ is therefore
given by
\begin{equation}
\label{A60KL}
A_{60} = {\cal K} - A_{61} \, \ln 2 + {\cal L}.
\end{equation}
Our improved results for $A_{60}$ coefficients rely essentially
on a more general code for the analytic calculations, written
in the computer-algebra package 
{\em Mathematica}~\cite{Wo1988,Disclaimer}, 
which enables the corrections to be evaluated 
semi-automatically. Intermediate expressions with some 
200,000 terms are encountered, and the complexity of the 
calculations sharply increases with the principal quantum number 
$n$, and, as far as the complexity of
the angular momentum algebra is concerned,
with the orbital angular quantum number of the bound electron. 

Of crucial importance was the development of convergence acceleration
methods which were used extensively for the evaluation of remaining
one-dimensional integrals which could not be done analytically.
These integrals are analogous to expressions encountered in
previous work (see~\cite[Eqs.~(36), (47) and~(48)]{jentschura96} and 
\cite[Eqs.~(80)--(84)]{JeSoMo1997}).
The numerically evaluated contributions
involve slowly convergent hypergeometric series,
and---in more extreme cases---infinite series 
over partial derivatives of hypergeometric functions, and
generalizations of Lerch's $\Phi$ 
transcendent~\cite{Ol1974,Ba1953vol1}. 
As a result of the summation over $l'$ in (\ref{SCGreensf}),
after performing radial integrals,
two specific hypergeometric functions enter naturally into 
the expressions for the bound-state matrix elements that 
characterize the one-loop correction (see, e.g.,~\cite[Eqs~(80)
and~(81)]{JeSoMo1997}). One of these
functions is given by
\begin{equation}
\label{Phi1}
\Phi_1(n,t) = {_2}F_1\left(1, -n t, 1 - nt, 
\left(\frac{1-t}{1+t}\right)^2\right),
\end{equation}
where the integration variable $t$ is in the range 0--1, 
and $n$ is the bound-state principal quantum number (${_2}F_1$ denotes the hypergeometric function---see, e.g., Chap.~15 in Ref.~\cite{abramovitz72}).
For $ t \simeq 0$, the power series expansion of $\Phi_1$
is slowly convergent,
\begin{equation}
\label{phi1power}
\Phi_1(n,t) = (n \, t) \sum_{k=0}^{\infty} \,
\frac{\left(\frac{1-t}{1+t}\right)^{2k}}{n\, t - k}.
\end{equation}
The series is nonalternating. In order to accelerate the 
convergence in the range $t \in (0,0.05)$, we employ
the combined nonlinear-condensation 
transformation~\cite{JeMoSoWe1999,AkSaJeBeSoMo2003}.
The other hypergeometric function that occurs naturally
in our calculations is
\begin{equation}
\label{Phi2}
\Phi_2(n,t) = {_2}F_1\left(1, -n t, 1 - n t, 
-\left( \frac{1-t}{1+t} \right) \right),
\end{equation}
For $0 < t < 0.05$, we accelerate the convergence of the
alternating power series 
\begin{equation}
\label{phi2power}
\Phi_2(n,t) = (n \, t) \sum_{k=0}^{\infty} \,
\frac{\left(-\frac{1-t}{1+t}\right)^{k}}{n\, t - k}
\end{equation}
via the $\delta$ transformation~\cite[Eq.~(8.4-4)]{We1989}.
The convergence acceleration leads to a much more reliable 
evaluation of the remaining numerical integrals which contribute to 
$A_{60}$ (but cannot be expressed in closed analytic form).
As a by-product of our investigations, we obtained   through this (independent) method Bethe logarithms 
which are consistent with the precise results of Ref.~\cite{goldman2000}. Here, we restrict the accuracy to 24 figures and
give results for P states:
\begin{eqnarray}\label{eqs:betheLog}
& & \ln k_0(2 {\mathrm P}) = \nonumber\\\nonumber
& & \quad -0.030~016~708~630~212~902~443~676(1),
\\[2ex]
& & \ln k_0(3 {\mathrm P})  = \nonumber\\\nonumber
& & \quad -0.038~190~229~385~312~447~701~163(1),
\\[2ex]
& & \ln k_0(4 {\mathrm P})  = \\\nonumber
& & \quad -0.041~954~894~598~085~548~671~037(1),
\\[2ex]
& & \ln k_0(5 {\mathrm P})  = \nonumber\\\nonumber
& & \quad -0.044~034~695~591~877~795~070~318(1).
\end{eqnarray}
These results, which test the numerical methods that we employed, are in agreement with other recent
calculations~\cite{haywood85,drake90note,FoHi1993,goldman2000}.

\begin{table}
\caption{\label{tbl:a60P}
Self-energy coefficient $A_{60}$~\refPar{Asixty} for P~states [see Eq.~\refPar{defFLOL1}].
The quoted error is due to numerical integration.
As in previous 
calculations (see Refs.~\cite{jentschura96,JeSoMo1997}), certain remaining 
one-dimensional integrals 
involving (partial derivatives of) hypergeometric functions could only be
evaluated numerically.
}
\begin{ruledtabular}
\begin{tabular}{cll}
\multicolumn{1}{c}{$n$} &
\multicolumn{1}{c}{${\rm P}_{1/2}$ ($\kappa = 1$)} &
\multicolumn{1}{c}{${\rm P}_{3/2}$ ($\kappa = -2$)} \\
\hline
2 & $-0.998~904~402(1)$ &  $-0.503~373~465(1)$ \\
3 & $-1.148~189~956(1)$
   &   $-0.597~569~388(1)$ \\
4 & $-1.195~688~142(1)$ &  $-0.630~945~795(1)$ \\
 5 & $-1.216~224~512(1)$ & $-0.647~013~508(1)$ \\
6 & $ -1.226~702~391(1)$ & $-0.656~154~893(1)$\\
7 & $ -1.232~715~957(1)$ & $-0.662~027~568(1)$ \\
\end{tabular}
\end{ruledtabular}
\end{table}

\begin{table}
\caption{\label{tbl:a60D}
$A_{60}$ coefficients~\refPar{Asixty} for D~states.}
\begin{ruledtabular}
\begin{tabular}{cll}
\multicolumn{1}{c}{$n$} &
\multicolumn{1}{c}{${\rm D}_{3/2}$ ($\kappa = 2$)} &
\multicolumn{1}{c}{${\rm D}_{5/2}$ ($\kappa = -3$)} \\
\hline
3 &  $0.005~551~573(1)$ & $0.027~609~989(1)$ \\
4 &   $0.005~585~985(1)$ & 
$0.031~411~862(1)$ \\
 5 & $0.006~152~175(1)$ & 
 $0.033~077~570(1)$ \\
6 &  $0.006~749~745(1)$ & $0.033~908~493(1)$ \\
7 & $0.007~277~403(1)$  & $0.034~355~926(1)$ \\
8 & $0.007~723~850(1)$ & $0.034~607~492(1)$ \\
\end{tabular}
\end{ruledtabular}
\end{table}

\begin{table}
\caption{\label{tbl:a60F}
$A_{60}$ coefficients~\refPar{Asixty} for F~states.}
\begin{ruledtabular}
\begin{tabular}{cll}
\multicolumn{1}{c}{ $n$} &
\multicolumn{1}{c}{${\rm F}_{5/2}$ ($\kappa = 3$)} &
\multicolumn{1}{c}{${\rm F}_{7/2}$ ($\kappa = -4$)} \\
\hline
4 &   $0.002~326~988(1)$ &   $0.007~074~961(1)$  \\ 
 5 & $0.002~403~158(1)$ &    $0.008~087~020(1)$ \\
\end{tabular}
\end{ruledtabular}
\end{table}

\begin{table}
\caption{\label{tbl:a60G}
$A_{60}$ coefficients~\refPar{Asixty} for G~states.}
\begin{ruledtabular}
\begin{tabular}{cll}
\multicolumn{1}{c}{$n$} &
\multicolumn{1}{c}{${\rm G}_{7/2}$ ($\kappa = 4$)} &
\multicolumn{1}{c}{${\rm G}_{9/2}$ ($\kappa = -5$)} \\
\hline
 5      & $0.000~814~415(1)$  &  $0.002~412~929(1)$ \\
\end{tabular}
\end{ruledtabular}
\end{table}

\begin{table}
\caption{\label{tableCKL}According to Eqs.~(\ref{HighGen}) and 
(\ref{LowGen}), the high- and low-energy parts can be cast into a general
form involving the terms ${\cal C}$, ${\cal K}$ and ${\cal L}$. 
The coefficient $A_{60}$ can be expressed in terms of ${\cal K}$,
$A_{61}$  and 
${\cal L}$ according to (\ref{A60KL}). Here, we present
analytic results
for the terms ${\cal C}$, $A_{61}$ and ${\cal K}$,  and numerical results for ${\cal L}$ 
(for states with $n=5$). The results for $A_{61}$ can be inferred  from 
Eqs.~(\ref{A61gen})--(\ref{A61L2}). For $l \geq 2$, we observe that
the $A_{61}$ are spin-independent and that ${\cal C} = A_{61}$.}
\begin{ruledtabular}
\begin{tabular}{ccccD{.}{.}{15}}
\multicolumn{5}{c}{\rule[-3mm]{0mm}{8mm} ${\cal C}$,
${\cal K}$ and ${\cal L}$ coefficients for states with $n=5$} \\
\multicolumn{1}{c}{\rule[-3mm]{0mm}{8mm} state} &
\multicolumn{1}{c}{\rule[-3mm]{0mm}{8mm} ${\cal C}$} &
\multicolumn{1}{c}{\rule[-3mm]{0mm}{8mm} $A_{61}$} &
\multicolumn{1}{c}{\rule[-3mm]{0mm}{8mm} ${\cal K}$} &
\multicolumn{1}{c}{\rule[-3mm]{0mm}{8mm} ${\cal L}$} \\
\hline
\rule[-3mm]{0mm}{8mm}
$5{\rm P}_{1/2}$ & 
$\frac{292}{1125}$ &
$\frac{796}{1125}$ &
$\frac{20129}{67500}$ & 
-1.023~991~781(1) \\
\rule[-3mm]{0mm}{8mm}
$5{\rm P}_{3/2}$ & 
$\frac{292}{1125}$ &
$\frac{436}{1125}$ &
$\frac{199387}{540000}$ & 
-0.747~615~653(1) \\
\rule[-3mm]{0mm}{8mm}
$5{\rm D}_{3/2}$ & 
$\frac{92}{7875}$ &
$\frac{92}{7875}$ &
$-\frac{35947}{3780000}$ & 
0.023~759~683(1) \\
\rule[-3mm]{0mm}{8mm}
$5{\rm D}_{5/2}$ & 
$\frac{92}{7875}$ &
$\frac{92}{7875}$ &
$\frac{3097}{157500}$ & 
0.021~511~798(1) \\
\rule[-3mm]{0mm}{8mm}
$5{\rm F}_{5/2}$ & 
$\frac{2}{1125}$ &
$\frac{2}{1125}$ &
$-\frac{2657}{1102500}$ & 
0.006~045~397(1) \\
\rule[-3mm]{0mm}{8mm}
$5{\rm F}_{7/2}$ & 
$\frac{2}{1125}$ &
$\frac{2}{1125}$ &
$\frac{774121}{211680000}$ & 
0.005~662~248(1) \\
\rule[-3mm]{0mm}{8mm}
$5{\rm G}_{7/2}$ & 
$\frac{2}{4725}$ &
$\frac{2}{4725}$ &
$-\frac{4397}{6048000}$ & 
0.001~834~827(1) \\
\rule[-3mm]{0mm}{8mm}
$5{\rm G}_{9/2}$ & 
$\frac{2}{4725}$ &
$\frac{2}{4725}$ &
$\frac{269}{283500}$ & 
0.001~757~471(1) \\
\end{tabular}
\end{ruledtabular}
\end{table}

The main results of this paper concerning the $A_{60}$ coefficients
are given in Tables~\ref{tbl:a60P}--\ref{tbl:a60G}, with an absolute
precision of $10^{-9}$.
In addition, we give explicit expressions for
the low- and high-energy parts of the self energy, for the states with
$n=5$ under investigation [see Eqs.~(\ref{HighGen}) and~(\ref{LowGen})
and Table~\ref{tableCKL}].  They may be helpful in an independent
verification of our calculations. Note that the ${\rm G}_{7/2}$ and
${\rm G}_{9/2}$ states involve the most problematic angular momentum
algebra of all atomic states considered here.

For some P states (see Table~\ref{tbl:a60P}), the values of $A_{60}$
reported here are four orders of magnitude more accurate than previous
results~\cite{jentschura96,JeSoMo1997}, due to the improved numerical
algorithms.  For the 3P$_{1/2}$ states, the numerical value for the
$A_{60}$ coefficients of Table~\ref{tbl:a60P} differs from the
previously reported result~\cite{JeSoMo1997} by more than the
numerical uncertainty quoted in Ref.~\cite{JeSoMo1997}, whereas they
are in agreement with previous results~\cite{jentschura96,JeSoMo1997}
in the case of 2P$_{1/2}$ and 4P$_{1/2}$ states.  The discrepancy for
$A_{60}(3\text{P}_{1/2})$ is on the level of $5 \times 10^{-4}$ in
absolute units, which corresponds to roughly 2~Hz (in frequency units) on the self-energy correction in
atomic hydrogen. The computational error in Ref.~\cite{JeSoMo1997}
was caused by numerical difficulties in one of the remaining
one-dimensional integrals involving the hypergeometric functions
(\ref{Phi1}) and (\ref{Phi2}), which could not be evaluated
analytically.  The numerical difficulties encountered in previous
calculations due to slow convergence of the integrals are essentially
removed by the convergence acceleration techniques.

\begin{table}
\caption{\label{cancel1}
As explained in Refs.~\cite{jentschura96,JeSoMo1997},
the low-energy contributions to $A_{60}$ 
naturally separate into the following
terms: the nonrelativistic quadrupole part 
$F_{\rm nq}$~\cite[Eq.~(39)]{jentschura96}, the
relativistic corrections to the current 
$F_{\rm \delta y}$~\cite[Eq.~(40)]{jentschura96},  
relativistic corrections to the Hamiltonian 
$F_{\rm \delta H}$~\cite[Eq.~(41)]{jentschura96},  
and relativistic corrections to 
the bound-state energy $F_{\rm \delta E}$~\cite[Eq.~(42)]{jentschura96} and
to the wavefunction $F_{\rm \delta \phi}$~\cite[Eq.~(43)]{jentschura96}.
This classification suggests that it is natural to refer
to the low-energy contribution ${\cal L}$ as a relativistic Bethe logarithm.
Observe that the total contribution to $A_{60}$ 
of the low-energy part, which reads
$0.001~834~827(1)$, is roughly five times smaller than
the largest individual contribution (from $F_{\rm \delta H}$),
due to mutual cancellations.}
\begin{ruledtabular}
\begin{tabular}{cD{.}{.}{15}}
\multicolumn{2}{c}{Contributions to the low-energy part ($5{\rm G}_{7/2}$)}\\
\hline
$A_{60}$-contribution due to $F_{\rm nq}$          & 
   0.002~875~830~9(5) \\
$A_{60}$-contribution due to $F_{\rm \delta y}$    & 
  -0.001~083~109~4(5) \\
$A_{60}$-contribution due to $F_{\rm \delta H}$    & 
  -0.008~917~782~1(5) \\
$A_{60}$-contribution due to $F_{\rm \delta E}$    & 
   0.004~920~556~0(5) \\
$A_{60}$-contribution due to $F_{\rm \delta \phi}$ & 
   0.004~039~332~1(5) \\
$A_{60}$ (see entry for ${\cal L}$ in 
Table~\ref{tableCKL}) 
&  0.001~834~827(1) \\
\end{tabular}
\end{ruledtabular}
\end{table}

\begin{table}
\caption{\label{cancel2}
For the $5{\rm G}_{7/2}$ state,
an additional numerical cancellation occurs when 
the finite contributions to $A_{60}$ originating
from the low-energy part (see the ninth row of
Table~\ref{tableCKL}) and the high-energy part
are added according to 
Eq.~(\ref{A60KL}). The high-energy 
contribution is $A_{60}(F_{\rm H}) = {\cal K} - A_{61}\,\ln 2$,
and the low-energy contribution is $A_{60}(F_{\rm L}) = {\cal L}$.}
\begin{ruledtabular}
\begin{tabular}{cd}
$A_{60}(F_{\rm H})$  & -0.001~020~413 \\
$A_{60}(F_{\rm L})$  &  0.001~834~828(1) \\
$A_{60}$       &  0.000~814~415(1) \\
\end{tabular}
\end{ruledtabular}
\end{table}

For some states, rather severe numerical
cancellations are observed between the high- and
low-energy contributions to the self energy, as well
as between the different contributions to the 
low-energy part.  This intriguing observation 
is documented in Tables~\ref{cancel1} and~\ref{cancel2},
using the $5{\rm G}_{7/2}$ state as an example.
Note that these numerical
cancellations go beyond the required exact, analytic
cancellation
of the divergent contributions which depend
on the scale-separation parameter $\epsilon$.

\typeout{Section: A60 for higher-n states}

\section{$A_{60}$ for higher-$n$ states}
\label{sec:a60higherN}

This section contains approximate formulas that we have found for the
$A_{60}$ coefficients of P and~D states, for principal quantum
numbers~$n$ that go beyond those of Tables \ref{tbl:a60P}
and~\ref{tbl:a60D}.  These tables contain enough values
of~$A_{60}(nl_j)$ for extrapolations to be made. We present the
asymptotic behavior of~$A_{60}(nl_j)$ as $n\rightarrow \infty$ as
\begin{subequations}
\label{eq:defApproxA}
\begin{equation}\label{eq:approxA60WithA3}
A_{60}(nl_j)
=
\approxA(n,l_j) + \landauO\left(\frac{1}{n^3}\right),
\end{equation}
where
\begin{equation}
\label{eq:defCoefsApproxA60}
\approxA(n,l_j)
\defi
\coefApproxLJ{0}
+
\frac{\coefApproxLJ{1}}{n}
+
\frac{\coefApproxLJ{2}}{n^2}.
\end{equation}
\end{subequations}
Such an asymptotic behavior is justified, for any non-S state, by its
similarity to the functional form of the self-energy
coefficient~$A_{61}$ in Eq.~\refPar{defFLOL1}---see
Eq.~\refPar{eq:A61partic}.  The values that we obtained for the
coefficients~$\coefApproxLJ{i}$ can be found in
Table~\ref{tbl:asymptCoefs}. The fitting method is described in
the Appendix.

\begin{figure*}
\includegraphics[width=0.8\linewidth]{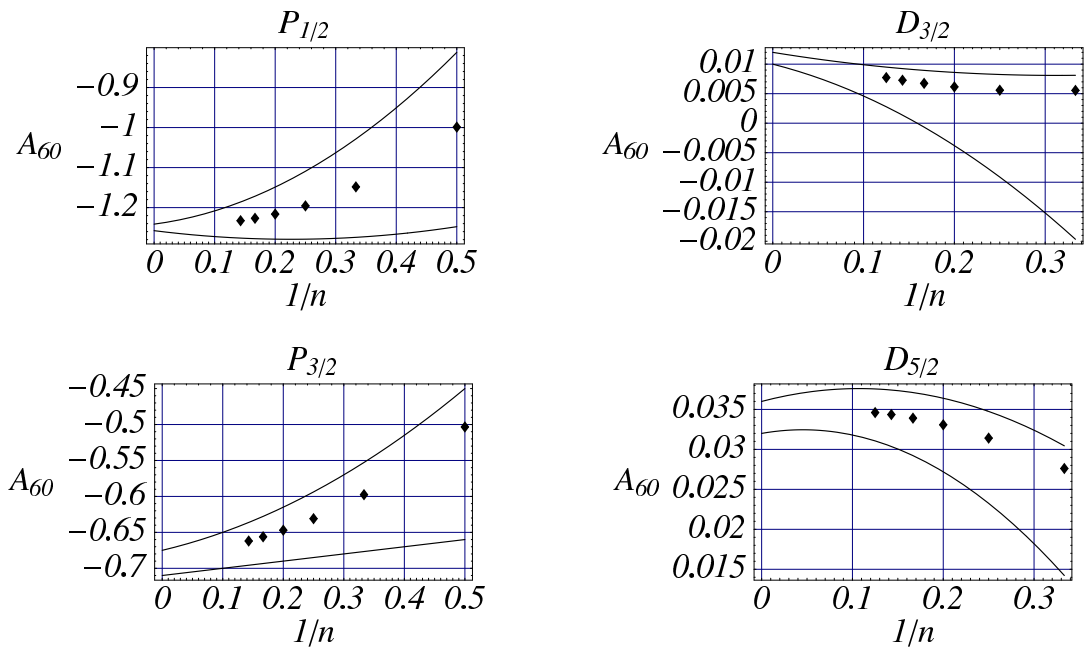}
\caption{\label{fig:approxA}
  These graphs show exact and approximate values of the self-energy
  coefficient~$A_{60}$---see Eq.~\refPar{defFLOL1}.  Exact values are
  represented by dots and can be found in Tables \ref{tbl:a60P}
  and~\ref{tbl:a60D}. The two curves of each graph represent the upper
  and lower limits of the approximation to~$A_{60}$ provided
  by~$\approxA$ in Eqs.~\refPar{eq:defApproxA}, by taking into account
  the uncertainties in the coefficients of
  Table~\ref{tbl:asymptCoefs}. For levels in hydrogen with principal
  quantum number $n\geq 10$, the uncertainty in~$A_{60}$ deduced from
  these curves contributes to the uncertainty in the electron self
  energy~\refPar{ESEasF} by less than 2~Hz. (The use of $1/n$ as the abscissa allows all large principal quantum numbers~$n$ to be represented in the graphs.)}
\end{figure*}

The approximation~$\approxA(n,l_j)$ to $A_{60}(nl_j)$ is depicted in
Fig.~\ref{fig:approxA}, for P and~D states.  According to the graphs in
this figure, the $\landauO(1/n^3)$ contribution
in~\refPar{eq:approxA60WithA3} is much smaller than the uncertainty
in~$\approxA$, which comes from the uncertainties in the coefficients
of Table~\ref{tbl:asymptCoefs}.

\begin{table}
\caption{\label{tbl:asymptCoefs}
The asymptotic behavior of~$A_{60}(nl_j)$ as $n\rightarrow\infty$ can be described by an expansion in~$1/n$. The following table contains the first  coefficients of such an expansion, as defined in Eq.~\refPar{eq:defApproxA}.
The approximate values of~$A_{60}(nl_j)$ that can be directly deduced from this table and from Eq.~\refPar{eq:defCoefsApproxA60} are the best available values of~$A_{60}$ for  P and D states, except for the states that are represented in Tables~\ref{tbl:a60P} and~\ref{tbl:a60D}. These results are depicted in Fig.~\ref{fig:approxA}.
}
\begin{ruledtabular}
\begin{tabular}{cddd}
state        
& \multicolumn{1}{c}{$\coefApprox{0}$}
& \multicolumn{1}{c}{$\coefApprox{1}$}
& \multicolumn{1}{c}{$\coefApprox{2}$}
 \\\hline
$\text{P}_{1/2}$ & -1.249(9) & 0.0(2) & 0.87(45) \\
$\text{P}_{3/2}$ & -0.69(2) & 0.15(5) & 0.25(25) \\
$\text{D}_{3/2}$ & 0.011(1) & -0.032(7) & -0.05(9) \\
$\text{D}_{5/2}$ & 0.034(2) & 0.025(5) & -0.18(4) \\
\end{tabular}
\end{ruledtabular}
\end{table}

The coefficients~$\coefApprox{i}$ of~\refPar{eq:defCoefsApproxA60}
given in Table~\ref{tbl:asymptCoefs} can be useful to spectroscopy
experiments that involve electronic levels with principal quantum
numbers that are higher than those of Tables \ref{tbl:a60P}
and~\ref{tbl:a60D}. In fact, the self energy of the electron of an
hydrogenlike ion can be estimated through Eqs.~\refPar{defFLOL1},
\refPar{A40gen}, \refPar{eq:A61partic} and~\refPar{eq:defApproxA},
with~$\approxA$ defined with the values of
Table~\ref{tbl:asymptCoefs}. Hydrogen has been and will be the subject
of extremely precise spectroscopy experiments, which now reach the
level of 1~Hz of uncertainty in transition frequencies.  The
uncertainty in the self energy~\refPar{ESEasF} that comes from the
uncertainties in the coefficients of Table~\ref{tbl:asymptCoefs}
through~\refPar{defFLOL1} and~\refPar{eq:defApproxA} is comparable to the
current experimental limit.  In fact, the uncertainties in~$\approxA$
in~\refPar{eq:defCoefsApproxA60} contribute to the self energy by less
than $\pm$2~Hz for $\text{P}_{1/2}$ states with $n>7$, less than
$\pm$1.6~Hz for $\text{P}_{3/2}$ states with $n>7$, less than
$\pm$0.12~Hz for $\text{D}_{5/2}$ states with $n>8$, and less than
$\pm$0.12~Hz for $\text{D}_{5/2}$ states with $n>8$ (precise values of~$A_{60}$ for lower values of~$n$ can be found in Tables \ref{tbl:a60P} and~\ref{tbl:a60D}).

Moreover, the coefficients of Table~\ref{tbl:asymptCoefs} can be
useful to theoretical calculations.  In fact, future values
of~$A_{60}$ for P and D states can be checked against the estimates
provided by~$\approxA$ in~\refPar{eq:defCoefsApproxA60}---see also the
curves of Fig.~\ref{fig:approxA}.

\section{Approximations of $A_{60}$ and of the Bethe logarithm}
\label{sec:approxA60andBL}
\typeout{Section: Approximation of A60 and of the Bethe logarithm}

In addition to studying the dependence of~$A_{60}(nl_j)$ on~$n$, as we
did in the previous section for P and~D states, it is interesting to
analyze the behavior of~$A_{60}(nl_j)$ as a function of~$l$, for
$j=l-1/2$ and $j=l+1/2$. We conjecture that $A_{60}(\bar{n} l_j)$, for
$\bar{n} \defi l+1$ and $j=l\pm 1/2$, decreases as
\begin{equation}\label{eq:asymptA60}
A_{60}(\bar{n} l_j)
\underset{l\rightarrow \infty}{\sim}
{\displaystyle\frac{c(j-l)}{l^k}}
\quad\text{with $k\geq 3$}
,
\end{equation}
where we probably have $k=4$ or $k=5$ [$c(1/2)$ and $c(-1/2)$ are two
unspecified numbers].  The form~\refPar{eq:asymptA60} is motivated in
this section.

We have also studied the asymptotic behavior of the Bethe logarithm
$\ln k_0(\bar{n}l) $, because this is a quantity similar to the
``relativistic Bethe logarithm''~$A_{60}$, and because it yields a
large contribution to the self energy [see Eqs.~\refPar{defFLOL1}
and~\refPar{A40gen}]. We show in this section that the Bethe logarithm
$\ln k_0(\bar{n}l) $, where $\bar{n}=l+1$, appears to
behave asymptotically as~$l^{-3}$.  This result differs from the~$l^{-7/2}$ asymptotic
behavior of $\ln k_0(\bar{n}l) $ deduced from Eq.~(B5)
in~\cite[p.~845]{erickson77}. Extrapolations of the Bethe logarithm
$\ln k_0(nl)$ as a function of~$n$ have been obtained through the method described in the Appendix, and used in
Ref.~\cite{kotochigova2002} for S, P and D states ($l=0$ to~$2$).

\begin{figure}
\includegraphics[width=0.7\linewidth]{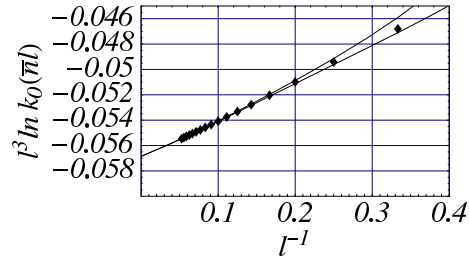}
\caption{\label{fig:fitBetheLog}
  Comparison between exact values of $l^3 \times \ln
  k_0(\bar{n}l) $ (dots) and the truncated asymptotic
  expansion of Eq.~\refPar{eq:blCoefs} (zone between the two
  curves)---$\ln k_0(\bar{n}l) $ is the Bethe logarithm,
  and $\bar{n} \defi l+1$.  The numerical values of the Bethe
  logarithms used in this graph~\cite{drake90note} are compatible with the
  values deduced from Eq.~\refPar{eq:blCoefs}, which are in the
  area between the two curves.  The fact that the data points seem to
  converge toward a finite value ($\simeq -0.057$) as~$l^{-1}\rightarrow
  0$ supports the conjecture of a~$l^{-3}$ asymptotic behavior of the
  Bethe logarithm $\ln k_0(\bar{n}l) $.}
\end{figure}

We also postulate that the Bethe logarithm $\ln k_0(\bar{n}l)$, where
$\bar{n}=l+1$, can be expanded in powers of $l^{-1}$ about~$l=\infty$.
In order to find the first five coefficients of such an expansion, we
used the fitting procedure described in the Appendix. The resulting
approximation reads:
\begin{eqnarray}\label{eq:blCoefs}
\lefteqn{l^3 \times  \ln k_0(\bar{n}l) 
\simeq
}&&
\\\nonumber
&&
\Bigg(
-0.056853(2)
\displaystyle
+
\frac{0.02478(4)}{l}
+
\frac{0.0387(8)}{l^2}
\\\nonumber
&&
\quad
{}
+
\frac{-0.114(6)}{l^3}
+
\frac{0.16(2)}{l^4}
\Bigg)
,
\end{eqnarray}
where $\bar{n}=l+1$, and where the neglected contribution is of
order~$l^{-5}$.  This approximation should be valid for~$l\rightarrow
\infty$; nevertheless, it yields values of the Bethe logarithm that
are both precise (see Fig.~\ref{fig:fitBetheLog}) and compatible with
all the values of $\ln k_0(\bar{n}l) $ for $l=3, \ldots,
19$ (taken from Ref.~\cite{drake90note}). For the $l\geq 20$ levels of
hydrogen, the uncertainty in the result of
approximation~\refPar{eq:blCoefs} is negligible, when compared to the
best experimental uncertainty in transition frequency measurements (about
1~Hz~\cite{biraben2001}).

Moreover, we suggest that the orders of magnitude of the self-energy
coefficient $A_{60}(nl_j)$ and of the Bethe logarithm $\ln k_0(nl)$ do
not depend on the principal quantum number~$n$, i.e., the order of
magnitude of a coefficient~$A_{60}(nl_j)$ is given by the order of
magnitude of~$A_{60}(\bar{n}l_j)$, where $\bar{n}=l+1$ (and similarly
for the Bethe logarithm).  For $A_{60}$, this behavior is a
generalization of what is observed for P, D, F and G states in Tables
\ref{tbl:a60P}--\ref{tbl:a60G}.  For the Bethe logarithm, the fact
that $\ln k_0(nl)$ and $\ln k_0(\bar{n}l)$ have the same order of
magnitude can be observed for states with $l < n \leq 20$ by
inspecting the results of Ref.~\cite{drake90note}.

The expressions \refPar{eq:asymptA60} and~\refPar{eq:blCoefs} for the
asymptotic behavior of~$A_{60}(\bar{n}l_j)$ and $\ln
k_0(\bar{n}l) $, where $\bar{n}=l+1$, could thus be used
for estimating the order of magnitude of the self
energy~\refPar{ESEasF}---with the help of Eqs.~\refPar{defFLOL1},
\refPar{A40gen}, \refPar{eq:A61partic}.  Estimating the self energy
correction~\refPar{ESEasF} can be useful in high-precision
spectroscopy experiments with large-$l$ levels.  Thus, for instance, a
recent experiment~\cite{devries2002} required evaluating the self
energies of circular ($n=l+1$) states of orbital quantum number
$l\simeq 30$. On the theoretical side, future calculations
of~$A_{60}(nl_j)$ and $\ln k_0(nl) $ can be checked
against the asymptotic behaviors of~$A_{60}(\bar{n}l_j)$ and $\ln
k_0(\bar{n}l) $ that are described above.

Since the order of magnitude of $A_{60}(n l_j)$ does not appear to
depend on~$n$, it is natural to represent it (for fixed $l$ and~$j$)
by the order of magnitude of either $\lim_{n\rightarrow \infty}
A_{60}(n l_j)$---largest possible~$n$--- or $A_{60}(\bar{n}
l_j)$---where $\bar{n}=l+1$ is the smallest~$n$ possible for the
angular momentum quantum number~$l$.  We chose the latter possibility
for two reasons. First, small-$n$ values of $A_{60}(n l_j)$ are
available (see Tables \ref{tbl:a60P}--\ref{tbl:a60G}).  Second, future
values of $A_{60}(n l_j)$ for higher angular quantum numbers~$l$ are
likely to be obtained first for states where $n=l+1$, which is the
smallest~$n$ possible for a given angular momentum quantum number~$l$.
In particular, such states have simpler radial wavefunctions (the
number of terms in the radial wavefunction of a state increases with
$n-l$).
And finally, circular states ($n=l+1$) are relevant to high-precision
spectroscopy experiments (see, e.g., Ref.~\cite{devries2002}), whereas
$n=\infty$ states are unphysical.

As mentioned above, we expect an asymptotic behavior of the form
$l^{-k}$, with~$k$ integer, for $A_{60}(\bar{n}l_j)$ and for the Bethe
logarithm $\ln k_0(\bar{n}l) $. Such a functional form is
motivated by the fact that all the~$A_{ik}(nl_j)$ coefficients of the
self-energy function~$F$ in Eq.~\refPar{defFLO} can be expanded in
power series of $1/n$ and~$l^{-1}$, except maybe for the two coefficients
related to this section, $A_{60}$ and~$A_{40}$, where the latter is a
function of the Bethe logarithm [see Eq.~\refPar{A40gen}].  (We
suppose that $A_{60}$ and~$A_{40}$ can also be expanded in such a
series.)
This can for instance be checked with the formulas for~$A_{ik}(nl_j)$
reviewed in Ref.~\cite[p.~468]{mohr2000b}, with the help of
Eq.~\refPar{A61gen} for~$A_{61}(nl_j)$, where $\Psi(n+1)$ can be
expanded in powers of $1/(n+1)$~\cite[\S~6.3.18]{abramovitz72}.

\begin{figure}
\includegraphics[width=0.7\linewidth]{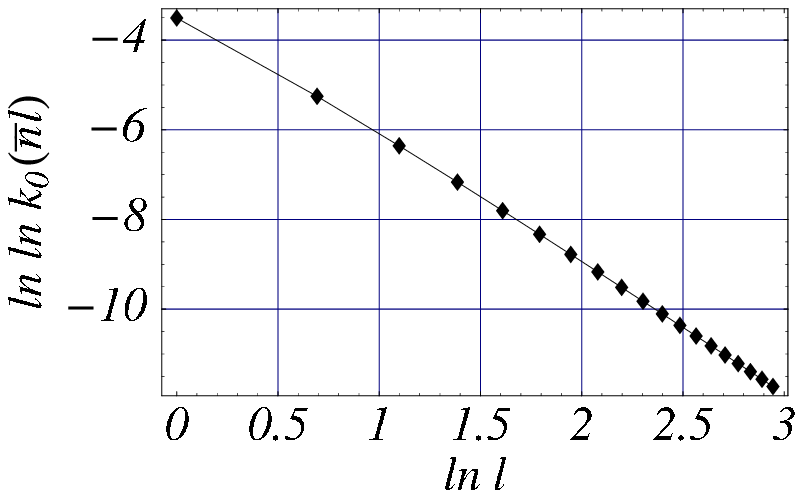}
\hspace*{0.2cm}\includegraphics[width=0.8\linewidth]{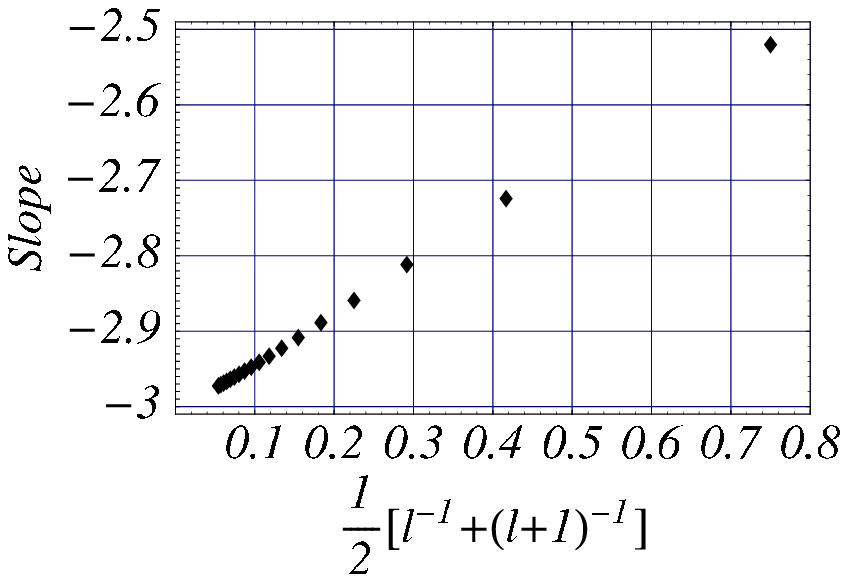}
\caption{\label{fig:logLogBethe}
  Upper graph: log-log plot of the Bethe logarithm $\ln
  k_0(\bar{n}l) $, where $\bar{n}\defi l+1$. Lower graph:
  slope between two successive points of the log-log plot.  The limit
  slope of~$-3$ as~$l\rightarrow \infty$ observed in the lower graph
  indicates that the Bethe logarithm $\ln k_0(\bar{n}l) $
  behaves asymptotically as~$l^{-3}$. This confirms what is observed in
  Fig.~\ref{fig:fitBetheLog}. (The abscissa of the points in the lower graph is chosen so as to produce a graph from which the limit slope of the upper graph as~$l\rightarrow\infty$ can be easily deduced.)}
\end{figure}

The $l^{-3}$ behavior of the Bethe logarithm $\ln
k_0(\bar{n}l) $, where $\bar{n}=l+1$, is suggested by
Fig.~\ref{fig:fitBetheLog}.  The points of this graph, which  represent 
\begin{equation}\label{eq:blMultipliedByL3}
l^3 \times \ln k_0(\bar{n}l) ,
\end{equation}
appear to converge toward a limit ($\simeq -0.057$) as $l^{-1}\rightarrow
0$.  We checked the~$l^{-3}$ behavior deduced from the study of
Eq.~\refPar{eq:blMultipliedByL3} by calculating the slope of a log-log
plot of the Bethe logarithm $\ln k_0(\bar{n}l) $ (with
numerical values taken from Ref.~\cite{drake90note}). The result, shown in
Fig.~\ref{fig:logLogBethe}, indicates that the Bethe logarithm does
indeed behave asymptotically as~$l^{-3}$; this coincides with the
conclusion from Fig.~\ref{fig:fitBetheLog}.

\begin{figure}
\includegraphics[width=0.7\linewidth]{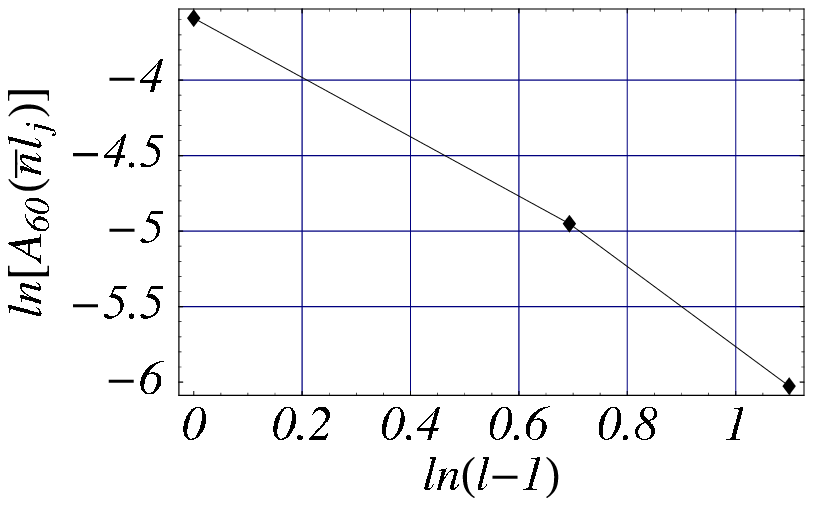}
\hspace*{0.32cm}\includegraphics[width=0.81\linewidth]{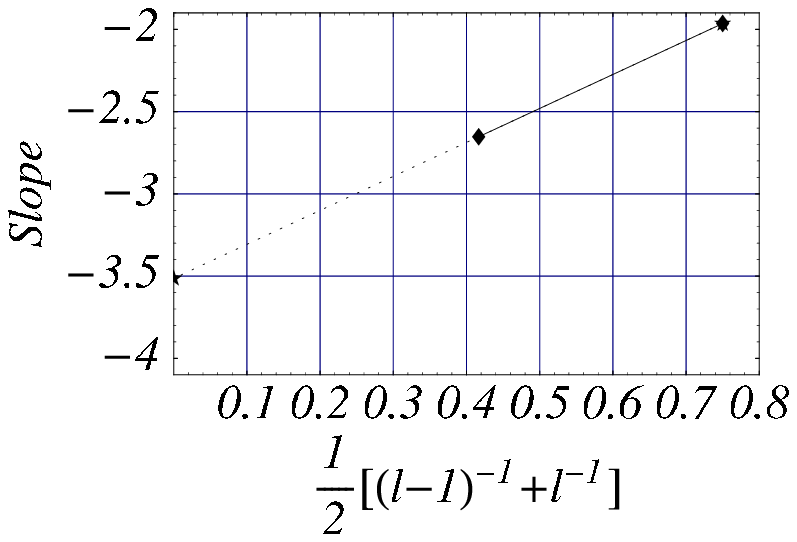}
\caption{\label{fig:slopeA60negKappa}
  Upper graph: log-log plot of the self-energy
  coefficient~$A_{60}(\bar{n}l_j)$, where $\bar{n}\defi l+1$ and
  $j=l+1/2$. Lower graph: slope between two successive points of the
  log-log plot (solid line) and extrapolation to~$l\rightarrow\infty$
  (dashes). By analogy with the graphs similarly obtained for the
  Bethe logarithm in Fig.~\ref{fig:logLogBethe}, we conclude that for
  $j=l+1/2$, $A_{60}(\bar{n}l_j)$ behaves asymptotically as $l^{-k}$
  with $k\geq 3$ and, probably, $k=4$ or $k=5$.  The values
  of~$A_{60}$ are taken from Tables~\ref{tbl:a60D}--\ref{tbl:a60G}.}
\end{figure}

\begin{figure}
\includegraphics[width=0.7\linewidth]{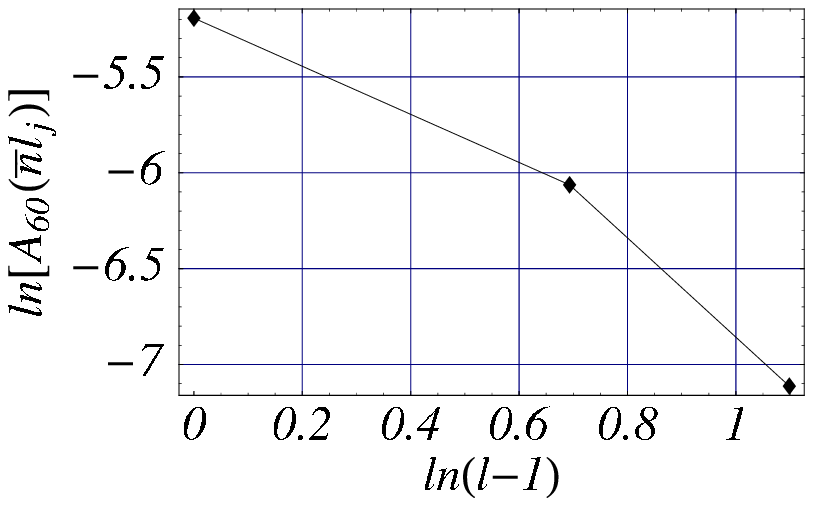}
\hspace*{0.55cm}\includegraphics[width=0.83\linewidth]{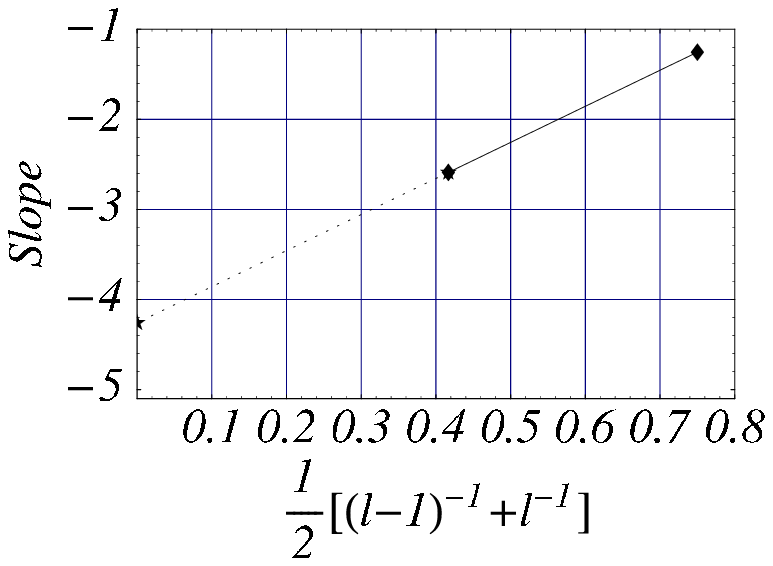}
\caption{\label{fig:slopeA60posKappa}
  Upper graph: log-log plot of the self-energy
  coefficient~$A_{60}(\bar{n}l_j)$, where $\bar{n}\defi l+1$ and
  $j=l-1/2$.  Lower graph: slope between two successive points of the
  log-log plot (solid line) and extrapolation to~$l\rightarrow\infty$
  (dashes). By analogy with the graphs similarly obtained for the
  Bethe logarithm in Fig.~\ref{fig:logLogBethe}, we conclude that for
  $j=l-1/2$, $A_{60}(\bar{n}l_j)$ behaves asymptotically as $l^{-k}$
  with $k\geq 3$ and, probably, $k=4$ or $k=5$.  The values
  of~$A_{60}$ are taken from Tables~\ref{tbl:a60D}--\ref{tbl:a60G}.  }
\end{figure}

\begin{figure}
\includegraphics[width=0.7\linewidth]{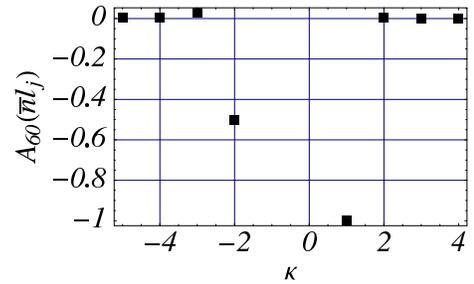}
\caption{\label{fig:ordersOfMagnitudeA60}
  This graph shows values of the self-energy
  coefficients~$A_{60}(\bar{n} l_j)$, where $\bar{n}\defi l+1$, as a
  function of the Dirac quantum number~$\kappa$, where~$\kappa$ is
  defined in~\refPar{eqs:defKappa}. The large value $A_{60}(1S_{1/2})
  \simeq -31$ is not represented here.
  This plot shows that for S and P states ($\kappa=-2$, $-1$ and~$1$),
  the $A_{60}$ coefficient exhibits an exceptional behavior; such an
  exceptional behavior is also found in the self-energy
  coefficient~$A_{61}$ in Eq.~\refPar{A61gen}, which is known
  analytically.}
\end{figure}

It is possible to use the procedure depicted in
Fig.~\ref{fig:logLogBethe} to estimate the integer exponent~$k$ of an
asymptotic behavior~$l^{-k}$ for the \emph{relativistic} Bethe logarithm
$A_{60}(\bar{n}l_j)$, where $\bar{n}=l+1$ and $j=l\pm 1/2$.  In fact,
it is reasonable to use the Bethe logarithm~$\ln
k_0(\bar{n}l) $ as a guide for studying the
\emph{relativistic} Bethe logarithm~$A_{60}$. Thus, the procedure
depicted in Fig.~\ref{fig:logLogBethe} was applied to the self-energy
coefficient~$A_{60}(\bar{n}l_j)$; we obtained the asymptotic behavior
presented at the beginning of this section, and in particular in
Eq.~\refPar{eq:asymptA60}. The graphs supporting~\refPar{eq:asymptA60}
are given in Fig.~\ref{fig:slopeA60negKappa}
for with states with $j=l+1/2$, and in Fig.~\ref{fig:slopeA60posKappa}
for states with $j=l-1/2$.  
Each of these graphs uses only three values of~$A_{60}$ (D, F and G
states); even though this is a relatively small number of values
compared to the number of available values of the Bethe logarithm, the
behavior of the first few data points in Fig.~\ref{fig:logLogBethe}
justifies using only a few small-$l$ values in order to predict the
asymptotic behavior of~$A_{60}(\bar{n}l_j)$ for~$l\rightarrow \infty$.

The values of the $A_{60}$ coefficient of~S and P states were not
used in obtaining Eq.~\refPar{eq:asymptA60}, because it is convenient
to treat the orders of magnitude of the $A_{60}$~coefficient of these
states separately from the orders of magnitude of higher-$l$ states;
Fig.~\ref{fig:ordersOfMagnitudeA60} illustrates this point.  We note
that the self-energy coefficient~$A_{61}$ also exhibits an exceptional
behavior for~S and P states (see, e.g., Eq.~(4.4a)
in~\cite{ErYe1965a}).  As an additional consequence, estimating the
coefficient~$c$ of the asymptotic form of~$A_{60}$ in
Eq.~\refPar{eq:asymptA60} would require use of states with orbital
angular momentum quantum number~$l\geq 2$ (D, F, etc.).

The possible values of the exponent~$k$ in Eq.~\refPar{eq:asymptA60}
deduced from both the graphs of Fig.~\ref{fig:slopeA60negKappa} and of
Fig.~\ref{fig:slopeA60posKappa} are compatible with each other ($k\geq
3$ with, probably, $k=4$ or $k=5$). It is indeed expected that the
asymptotic form of~$A_{60}(\bar{n}l_j)$ be the same for $j=l+1/2$
and $j=l-1/2$, as can be seen from the numerical values for D, F and G
states found in Tables~\ref{tbl:a60D}--\ref{tbl:a60G}.  More precise
estimates of the asymptotic exponent~$k$ in Eq.~\refPar{eq:asymptA60}
can be obtained through the procedure we used in
Figs.~\ref{fig:slopeA60negKappa} and~\ref{fig:slopeA60posKappa}, as
soon as additional values of $A_{60}(\bar{n}l_j)$, with $\bar{n}\defi
l+1$ are available.

According to the results of this section, the ``\emph{relativistic}
Bethe logarithm''~$A_{60}(\bar{n}l_j)$ decreases at least as fast (and
probably one or two powers faster), as a function of~$l$, than the
Bethe logarithm~$\ln k_0(\bar{n}{l})$.  Such a behavior is also found
in the (Dirac-Coulomb) energy of hydrogen and hydrogenlike ions.
Thus, the Dirac-Coulomb energy of an electron bound to a nucleus of
charge number~$Z$ is (see, e.g., \cite[p.~466]{mohr2000b})
\begin{equation}\label{eq:diracEnergy}
E_{nj} = \left[ 1+ \frac{(Z\alpha)^2}{(n-\delta)^2}\right]^{-1/2},
\end{equation}
where
\[
\delta \defi (j+1/2) - \sqrt{(j+1/2)^2 - (Z\alpha)^2}.
\]
According to~\refPar{eq:diracEnergy}, an electron in a circular state
$\bar{n}l_j$ with $j=l+1/2$ (and $\bar{n}=l+1$) has an energy
\begin{equation}\label{eq:diracCircularNegKappa}
E_{\bar{n}, l+\frac{1}{2}}
=
\sqrt{
{1-[Z\alpha/(l+1)]^2}}.
\end{equation}
In the Taylor expansion (in $Z\alpha$) of this energy, the asymptotic
behavior of the coefficient of $(Z\alpha)^{2k}$ is given by~$l^{-2k}$
(this conclusion also holds for circular state $\bar{n}l_j$ with
$j=l-1/2$). Thus, for circular states, successive relativistic
corrections to the nonrelativistic energy of a bound electron fall off
faster and faster with the orbital quantum number~$l$,
with two additional powers of $l^{-1}$ for each order in~$(Z\alpha)^2$.
If this rule applies to the coefficients of the self-energy
expansion~\refPar{defFLOL1}, the asymptotic form of
$A_{60}(\bar{n}l_j)$ as $l\rightarrow \infty$ should be~$l^{-4}$; in
fact, the lower-order coefficient~$A_{40}(\bar{n}l_j)$ decreases
as~$l^{-2}$, as can be seen in Eq.~\refPar{A40gen}. On the other hand,
since $A_{60}(nl_j)$ can be considered as a relativistic correction to
the Bethe logarithm, applying the above rule yields an asymptotic form
in~$l^{-5}$ for $A_{60}(\bar{n}l_j)$, since the
Bethe logarithm behaves as~$l^{-3}$, as described in this section.
These observations are fully compatible with the graphs of
Figs.~\ref{fig:slopeA60negKappa} and~\ref{fig:slopeA60posKappa}, from
which the asymptotic form~\refPar{eq:asymptA60}
of~$A_{60}(\bar{n}l_j)$ was deduced (with an exponent~$k$ probably
equal to 4 or~5).

\typeout{Section: numerical tests}

\section{Checks of the $A_{60}$ coefficients}
\label{sec:check}

We have checked our analytic results for~$A_{60}$ (cf.\ 
Tables~\ref{tbl:a60P}--\ref{tbl:a60G}) by an independent method: the
analytic results were compared to values deduced from
non-perturbative, numerical calculations of the self
energy~\refPar{ESEasF}.  We have used the numerical self-energy values
of
Refs.~\cite{LBInMo2001,jentschura2001b,indelicato98,indelicato98a,mohr92,mohr92b},
as well as new values~\cite{lebigot2002b}, which extend the results of
Ref.~\cite{LBInMo2001} to smaller nuclear charge numbers~$Z$ (to $Z$
between 10 and 25). In most cases, the checks that we detail below
\emph{confirm} the values of~$A_{60}$ reported in
Tables~\ref{tbl:a60P}--\ref{tbl:a60G}, to a relative precision of
about~15~\%. The few exceptions are the following. For 2P states, the
numerical values of the self-energy confirm the results of
Table~\ref{tbl:a60P} to about~1~\%.  For $n\text{D}_{3/2}$ states with
$n=3,\ldots,8$, the non-perturbative self-energy results yield
$A_{60}(n\text{D}_{3/2})= 0.005(10)$, in agreement with the results of
Table~\ref{tbl:a60D}. And finally, we did not check
$A_{60}(8\text{D}_{5/2})$ in Table~\ref{tbl:a60D} by using
non-perturbative self-energy values because no such values are
available for the $8\text{D}_{5/2}$ state.  However, as depicted in
Fig.~\ref{fig:approxA}, the value of $A_{60}(8\text{D}_{5/2})$
reported here appears to fit well within the series of
$A_{60}(n\text{D}_{5/2})$ values for $n=3,\ldots,7$ (see
Table~\ref{tbl:a60D}).

\begin{figure}
\begin{flushleft}
(a)\hspace{0.55cm}\parbox{0.8\linewidth}{\includegraphics[width=\linewidth]{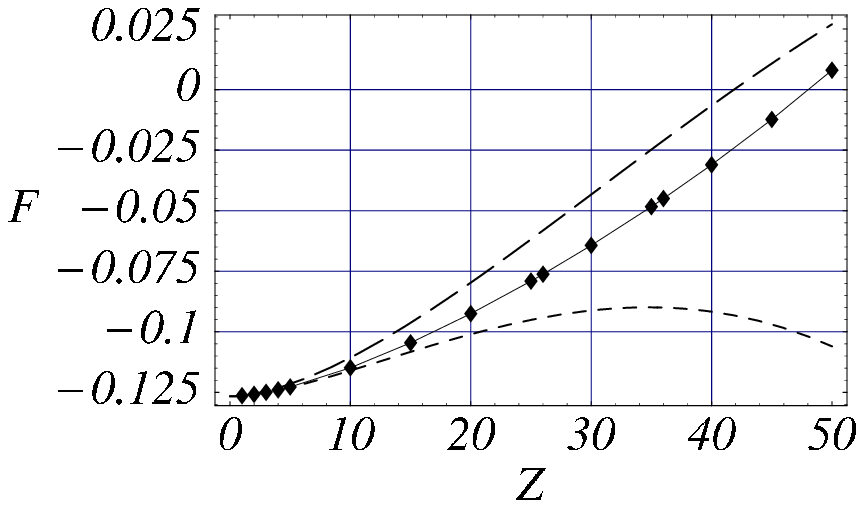}}\\[.25cm]
(b)\hspace*{0.90cm}\parbox{0.8\linewidth}{\includegraphics[width=\linewidth]{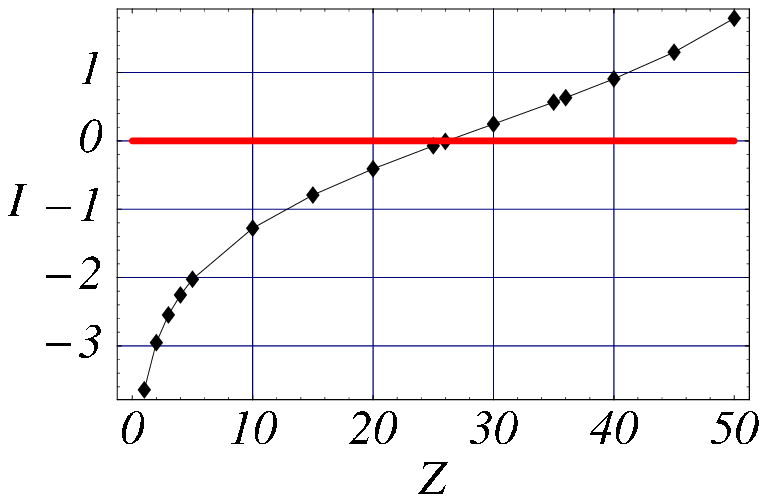}}
\end{flushleft}
\caption{\label{fig:2p1}
  Figure~(a) shows exact and approximate values of the (scaled) self
  energy~$F$ of a $2\text{P}_{1/2}$ electron [see
  Eq.~\refPar{ESEasF}]. Exact values are given on the solid line. The
  two-coefficient approximation~\refPar{approxOrder2} is represented
  by long dashes.  The three-coefficient
  approximation~\refPar{approxOrder3} uses the value
  of~$A_{60}(2\text{P}_{1/2})$ that we provide in
  Table~\ref{tbl:a60P}, and is indicated by short dashes. Figure~(b)
  displays the improvement provided by the inclusion of~$A_{60}$ in
  the self-energy approximation, as measured by the function~$I$ in
  Eq.~\refPar{improvementFunction}; negative values of~$I$ indicate
  that including~$A_{60}$ improves the approximation.}
\end{figure}

\begin{figure}
\begin{flushleft}
(a)\hspace{0.55cm}\parbox{0.8\linewidth}{\includegraphics[width=\linewidth]{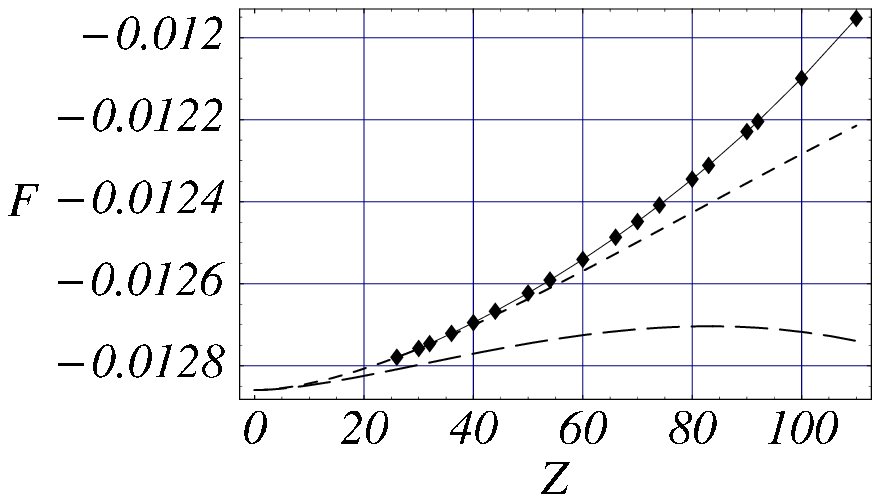}}\\[.25cm]
(b)\hspace*{1.0cm}\parbox{0.8\linewidth}{\includegraphics[width=\linewidth]{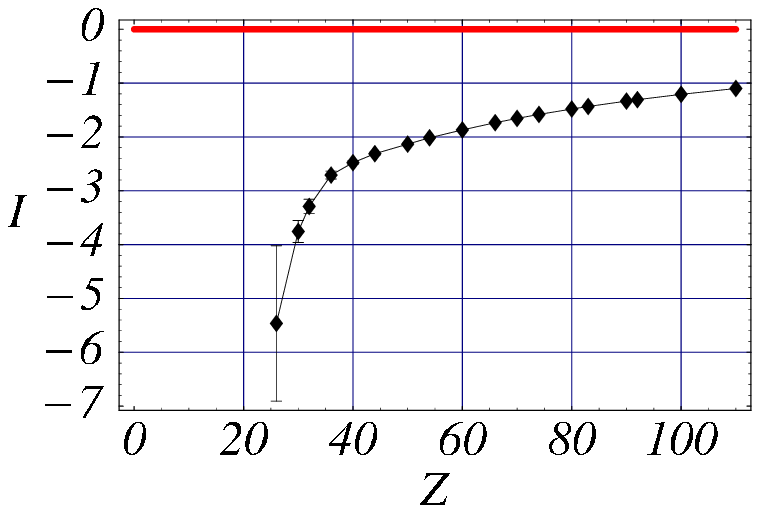}}
\end{flushleft}
\caption{\label{fig:5g7}
  These two figures represent respectively the same quantities as
  those found in Fig.~\ref{fig:2p1}, but for the $5\text{G}_{7/2}$
  level instead of the $2\text{P}_{1/2}$ level. 
  The fact that curve~(b) contains negative values of $I$ [see
  Eq.~\refPar{improvementFunction}] indicates that the three-order
  approximation~\refPar{approxOrder3} to the self
  energy~\refPar{defFLOL1} is better than the two-order
  approximation~\refPar{approxOrder2}, at least over the range of
  nuclear charge numbers $Z=25$--$110$. The three-order
  approximation~\refPar{approxOrder3} uses the value
  of~$A_{60}(5G_{7/2})$ reported in Table~\ref{tbl:a60G}.}
\end{figure}

The first check that we applied consisted
of comparing the numerical, exact results for~$F$ to 
two of its successive approximations. The first approximation,
$F^{(2)}(Z\alpha)$, includes the \emph{two} dominant and already-known
coefficients $A_{40}$~\refPar{A40gen} and $A_{61}$~\refPar{A61gen} of
expansion~\refPar{defFLOL1}:
\begin{equation}\label{approxOrder2}
F^{(2)}(Z\alpha) \defi A_{40} +  (Z\alpha)^2  A_{61} \ln(Z\alpha)^{-2},
\end{equation}
and the second approximation, $F^{(3)}$, includes in addition the
next-order contribution reported in this paper:
\begin{equation}\label{approxOrder3}
F^{(3)}(Z\alpha) \defi A_{40} +  (Z\alpha)^2  \left [A_{61} \ln(Z\alpha)^{-2}
+ A_{60} \right].
\end{equation}
For a given electronic level $nl_j$, one expects that for low~$Z$, the
curve of the higher-order approximation~$F^{(3)}(Z\alpha)$ be closer
to the curve of~$F(Z\alpha)$ than~$F^{(2)}(Z\alpha)$.  In order to
check this, we plotted the quantity
\begin{equation}\label{improvementFunction}
I(Z\alpha) 
\defi
\ln \left|
\frac{F(Z\alpha) - F^{(3)}(Z\alpha)
}{F(Z\alpha) - F^{(2)}(Z\alpha)}
\right|,
\end{equation}
which should go to~$-\infty$ as $Z\rightarrow 0$, as can be seen from
Eq.~(\ref{defFLOL1}).  In~\refPar{improvementFunction}, the purpose of
the logarithm is only to obtain more legible graphs; a value
of~$I$ lower than zero indicates that including~$A_{60}$ in the
approximation of~$F$ improves the lower-order approximation.  For the
states of Tables~\ref{tbl:a60P}--\ref{tbl:a60G}, graphs
of~\refPar{improvementFunction} are compatible with their expected
behavior [$I(Z\alpha)$ is negative for $Z$ sufficiently close to zero,
and is consistent with a~$-\infty$ limit].  Figures~\ref{fig:2p1}
and~\ref{fig:5g7} show this behavior for two electronic states.

Moreover, the improvement provided by the inclusion of~$A_{60}$ in the
approximation for~$F$ becomes greater as the total angular
momentum~$j$ increases: for given $n$ and~$Z$, the improvement
function~\refPar{improvementFunction} decreases as $j$~increases; this
behavior can observed by comparing Figs.~\ref{fig:2p1} and~\ref{fig:5g7}.
Similarily, the range of~$Z$ for which approximation~$F^{(3)}$ is
better than~$F^{(2)}$ increases with increasing~$j$.  In the worst of
the cases considered here ($j=1/2$), approximation~$F^{(3)}$ is better
than~$F^{(2)}$ up to $Z \simeq 25$.  As shown in Fig.~\ref{fig:5g7},
for a high-$j$ level such as $5\text{G}_{7/2}$, the higher-order
approximation~$F^{(3)}$ is better than~$F^{(2)}$ even up to~$Z=110$.

\begin{figure}
\includegraphics[width=0.7\linewidth]{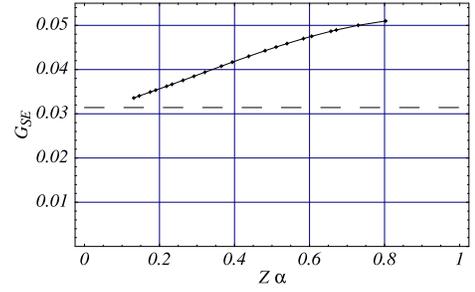}
\caption{\label{fig:GSE}
  Plot (solid line) of numerical values of the
  remainder~$G_{\text{SE}}(4\text{D}_{5/2}, Z\alpha)$ of the self
  energy~\refPar{defFLO}; the dashed line indicates the value
  of~$A_{60}(4\text{D}_{5/2})\simeq 0.0314$ reported in this paper (see
  Table~\ref{tbl:a60D}). By definition, the coefficient~$A_{60}$ can
  be obtained as the limit~\refPar{Asixty} of~$G_{\text{SE}}$ as
  $Z\alpha \rightarrow 0$. This plot shows that the value of~$A_{60}$
  extracted from numerical self-energies is consistent with the value
  obtained by the calculations presented in this paper. We made
  identical observations for all the states of
  Tables~\ref{tbl:a60P}--\ref{tbl:a60G}.}
\end{figure}

The second check consisted in estimating~$A_{60}$ from the numerical
values of the self energy~(\ref{ESEasF}).  For all the electronic
levels $nl_j$ studied here (except for~$8\text{D}_{5/2}$), we have plotted the
function~$G_{\text{SE}}(nl_j, Z\alpha)$ of~\refPar{defFLO}; this is
made possible by the fact that all the coefficients of~\refPar{defFLO}
are (analytically) known for any state~\cite{ErYe1965a,ErYe1965b}, except
for the Bethe logarithm, which has been numerically evaluated for many states, including
the ones we consider here~\cite{haywood85,drake90note,FoHi1993,goldman2000}.  As
indicated in~\refPar{Asixty}, the limit of the
remainder~$G_{\text{SE}}(nl_j, Z\alpha)$ as $Z\alpha \to 0$ is by
definition~$A_{60}(nl_j)$. We have estimated this limit both visually
and by fitting~$G_{\text{SE}}(nl_j, Z\alpha)$ with various choices of
non-zero higher-order terms.  A typical curve
for~$G_{\text{SE}}(Z\alpha)$ is shown in Fig.~\ref{fig:GSE}.  The
estimates of~$A_{60}$ obtained by these procedures \emph{confirm} the
independent analytic results of Tables~\ref{tbl:a60P}--\ref{tbl:a60G}
to a typical accuracy of~10--20~\%, with a few exceptions.  Thus, for
2P levels, plotting~$G_{\text{SE}}$ as in Fig.~\ref{fig:GSE} allowed
us to confirm the values of $A_{60}(2{\text{P}}_j)$ in
Table~\ref{tbl:a60P} to a precision of about 1~\%.
This higher precision is obtained by using the self
energies of 2P states obtained in Ref.~\cite{jentschura2001b} for values
of~$Z\alpha$ close to
zero ($Z=1,\ldots,5$): such low-$Z$ self energies are well-suited to an evaluation
of~$A_{60}$ by the limit~\refPar{Asixty}. Plotting~$G_{\text{SE}}$ for
$\text{D}_{3/2}$ states lead to $A_{60}(n\text{D}_{3/2})=
0.005(10)$ for $n=3,\ldots,8$, in agreement with Table~\ref{tbl:a60D}.
Finally, since no non-perturbative self-energy~\refPar{ESEasF} is
available for~$8\text{D}_{5/2}$ states, we were not able to
independently obtain $A_{60}(8\text{D}_{5/2})$ by using such values.

As a by-product of our work with graphs of $G_{\text{SE}}(nl_j,
Z\alpha)$, we estimate the self-energy remainder
$G_{\text{SE}}(nl_j,\alpha)$ relevant to hydrogen ($Z=1$) to be
0.030(5) for $3\text{D}_{5/2}$ and $4\text{D}_{5/2}$ states [see
Eq.~\refPar{defFLO}]; this is larger than the estimate of 0.00(1)
given in Ref.~\cite[p.~468]{mohr2000b}.  These two
new values change the previous estimate of the self energy of
$3\text{D}_{5/2}$ and $4\text{D}_{5/2}$ states through
Eq.~\refPar{defFLOL1} by a relatively large amount, compared to the
current best experimental uncertainty in transition frequencies (about
1~Hz~\cite{biraben2001}).  Thus, a variation of 0.03 in
$G_{\text{SE}}(3\text{D}_{5/2},\alpha)$ in~\refPar{defFLO} corresponds
to a variation of about 50~Hz in the self energy
correction~\refPar{ESEasF} of the $3\text{D}_{5/2}$ level in hydrogen.
The same variation in $G_{\text{SE}}(4\text{D}_{5/2},\alpha)$ induces
a variation of about 20~Hz in the self energy of the $4\text{D}_{5/2}$
level in hydrogen; on the other hand, this latter change is small
compared to the uncertainty of the relevant measurements considered in
Ref.~\cite{mohr2000b}.

\begin{figure}%
\vspace*{-0.3cm}\includegraphics[width=0.75\linewidth]{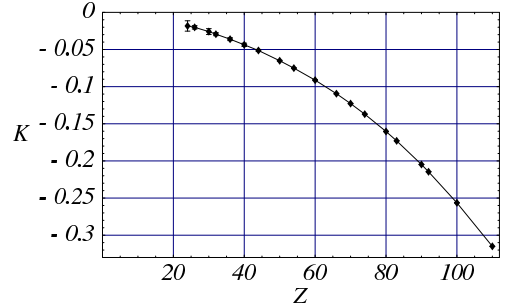}
\caption{\label{fig:deltaGSE}
  Plot of the function~$K$ in Eq.~\refPar{deltaA60check} for the
  $5\text{F}_{7/2}$ and~$5\text{F}_{5/2}$ states.  The limit of this
  function as $Z\rightarrow 0$ must be zero if the
  coefficients~$A_{60}$ of Tables~\ref{tbl:a60P}--\ref{tbl:a60G} agree
  with exact, numerical values of the self energy. 
  The curve displayed here indicates that the two values of $\DeltaFS
  A_{60}(5F)$ in Eq.~\refPar{defDeltaA60} obtained independently from
  Table~\ref{tbl:a60F} and from non-perturbative self
  energies~\refPar{ESEasF} [via Eqs.~\refPar{defFLOL1},
  \refPar{A40gen}, \refPar{A61L2}, \refPar{deltaGSE} and~\refPar{limDeltaGSE}] do not differ by more than about
  3~\%.}
\end{figure}

As a third and last check, we used the numerical, exact
values of~$F$ in order to study the following difference between
remainders~$G_{\text{SE}}$ [see Eqs.\ \refPar{defFLO} and
\refPar{defFLOL1}]:
\begin{equation}\label{deltaGSE}
\DeltaFS G_{\text{SE}}(nl, Z\alpha) \defi
G_{\text{SE}}(n l_{l+1/2}, Z \alpha) 
- G_{\text{SE}}(n l_{l-1/2}, Z \alpha),
\end{equation}
where, by definition of $A_{60}$~\refPar{Asixty},
\begin{eqnarray}\label{limDeltaGSE}
\lefteqn{\lim_{Z\alpha \rightarrow 0}
\DeltaFS G_{\text{SE}}(nl, Z\alpha)  }
\qquad &
\\\nonumber
&=&
A_{60}(n l_{l+1/2}) -
A_{60}(n l_{l-1/2})
\\\label{defDeltaA60}
&\defi&
\DeltaFS A_{60}(nl)\, ,
\end{eqnarray}
which denotes a quantity associated to the fine-structure.
The numerical evaluation of this limit is interesting: for the states
of Tables~\ref{tbl:a60P}--\ref{tbl:a60G}, the numerical results
for~$F$ yield values of $\DeltaFS A_{60}(nl)$ that are more accurate
than our numerical estimates of the two individual terms $A_{60}(n
l_{l+1/2})$ and $A_{60}(n l_{l-1/2})$.  Our analytic values for
$\DeltaFS A_{60}$ in Eq.~(\ref{defDeltaA60}) were
checked by plotting
\begin{equation}\label{deltaA60check}
K(Z) \defi  \frac{\DeltaFS G_{\text{SE}}(nl, Z\alpha)}{\DeltaFS A_{60}(nl)}
- 1,
\end{equation}
where~$\DeltaFS G_{\text{SE}}(nl, Z\alpha)$ was calculated from the
\emph{numerical} values of~$F$ [see Eq.~\refPar{defFLOL1} and the
coefficients reproduced in Sec.~\ref{notations}], and where the value
of~$\DeltaFS A_{60}(nl)$ in Eq.~\refPar{defDeltaA60} was deduced from
the \emph{analytic} results of
Tables~\ref{tbl:a60P}--\ref{tbl:a60G}\@.  If the numerical and
analytic estimates of $\DeltaFS A_{60}(nl)$ do agree, the
function~\refPar{deltaA60check} goes to zero as $Z \rightarrow 0$.
This is indeed consistent with what we observed;
figure~\ref{fig:deltaGSE} provides an example of this behavior. We
\emph{confirm} the values of $\DeltaFS A_{60}(nl)$ in
Eq.~\refPar{defDeltaA60} that can be immediately deduced from
Tables~\ref{tbl:a60P}--\ref{tbl:a60G}\@.  The analytic results for
$\DeltaFS A_{60}(nl)$ are thus found to be consistent with the
numerical data for~$\DeltaFS G_{\text{SE}}$; the level of confirmation is
5--10~\% [relative to $\DeltaFS A_{60}(nl)$] for P and D~states (1~\%
for the 2P states, and $8\text{D}$ states not included, for the reason
mentioned above), 3~\% for F states, and~1\% for G~states.

This represents an \emph{improvement} over the accuracy
of~$A_{60}(nl_j)$ obtained by the previous check. This improvement
comes evidently from the fact that the relative deviation of $\DeltaFS
G_{\text{SE}}$ in Eq.~\refPar{deltaGSE} from $\DeltaFS A_{60}$ in
Eq.~\refPar{defDeltaA60} is small over the whole range $ 0 < Z \leq
110 $, compared to the relative deviation
\begin{equation}\label{deviationA60}
\frac{ G_{\text{SE}}(nl_j, Z\alpha)}{ A_{60}(nl_j)}
- 1,
\end{equation}
of $G_{\text{SE}}$ [see Eq.~\refPar{defFLO}] from $A_{60}(nl_j)$ in
Eq.~\refPar{Asixty}---with $j= l + 1/2$ or $j=l-1/2$.  As a
consequence, the uncertainty in the numerical evaluation of the limit
of~\refPar{deltaA60check} as $Z \rightarrow 0$ is relatively small.
Figure~\ref{fig:deltaGSE} shows an example of the smallness of the
contributions to $\DeltaFS G_{\text{SE}}$ that go beyond $\DeltaFS
A_{60}$.  Moreover, we have observed that the higher the angular
momentum~$l$, the smaller the values of the
deviation~\refPar{deltaA60check}, hence the stronger confirmation of
our values of~$\DeltaFS A_{60}(nl)$ for high orbital angular momenta.

\typeout{Section: Summary of Results}
\section{Summary of Results}
\label{SummaryOfResults}

This paper contains results that are relevant to the self energy of a
non-S electron bound to a point nucleus of charge number~$Z$. We
provided estimates and values (see also Ref.~\cite{jentschura2003b})
for the first two non-analytically-known contributions to the
self-energy expansion~\refPar{defFLO}, namely the Bethe logarithm $\ln
k_0(nl)$ and the so-called $A_{60}(nl_j)$ coefficient, which can be
viewed as a \emph{relativistic} Bethe logarithm.  The main numerical
results are contained in Tables~\ref{tbl:a60P}--\ref{tbl:a60G}, in
Eq.~\refPar{eq:defApproxA} and Table~\ref{tbl:asymptCoefs}, in
Eq.~\refPar{eq:asymptA60} and in Eq.~\refPar{eq:blCoefs}.  We have
also conjectured, in Sec.~\ref{sec:approxA60andBL}, that the
relativistic Bethe logarithm~$A_{60}(nl_j)$ does not strongly depend
on the principal quantum number~$n$.  In addition to this, we note
that the orders of magnitude of $A_{60}(n\, l_{l-1/2})$ and
$A_{60}(n\, (l+1)_{l+3/2})$ are the same (for a given set of quantum numbers $n$ and~$l>1$), in
Tables~\ref{tbl:a60P}--\ref{tbl:a60G}.  These results, taken together, yield in
particular
the
best available approximations of the self energy in hydrogen and light
hydrogenlike ions, except for $n=1$ and $n=2$
levels~\cite{JeMoSo1999,jentschura2001b} (see also
Sec.~\ref{sec:check}); such an approximation can be obtained through
Eqs.~\refPar{ESEasF} and~\refPar{defFLOL1}.

Calculating $A_{60}$ has been a challenge since the seminal work of
Bethe~\cite{Be1947} on the dominant self-energy coefficients of
S~states [see Eqs.~\refPar{defFLOL1} and~\refPar{ESEasF}].  Details of
the method we used were described in Sec.~\ref{EpsilonMethod}
and~\ref{ResHighLow}.  As discussed in Sec.~\ref{sec:check}, including
the coefficients~$A_{60}$ reported in
Tables~\ref{tbl:a60P}--\ref{tbl:a60G} in a (truncated) expansion of
the self energy improves its accuracy over a large range of nuclear
charge numbers~$Z$.

We checked our calculations of~$A_{60}$ by both analytic and numerical
means. The so-called $\epsilon$ method, which we have employed (see
Sec.~\ref{EpsilonMethod}), makes divergences appear in the low- and
high-energy contributions to~$A_{60}$, as the scale-separating
parameter~$\epsilon$ between these two contributions goes to zero. We
have observed that, as required, these divergences cancel when the two
parts are added.  Moreover, our calculations correctly reproduced the
known lower-order coefficients $A_{40}$ and~$A_{61}$.  We have also
checked our results for~$A_{60}$ against numerical values of the self
energy, and were able to confirm them by this independent method to
the level of about~15~\% (except for $\text{D}_{3/2}$ states, as
explained in Sect.~\ref{sec:check}).

Obtaining results for $A_{60}$ required extending (analytically) the
angular algebra developed for 2P states~\cite{jentschura96} to higher
angular momenta.  Techniques of numerical convergence acceleration of
series~\cite{jentschura99d,JeMoSoWe1999,AkSaJeBeSoMo2003} were instrumental
in evaluating the parts of~$A_{60}$ that could not be analytically
calculated. The recent analytic calculations of Ref.~\cite{LBInMo2001}
enabled us to obtain with a high precision the
self energy~\refPar{ESEasF} of electrons with high ($j > 3/2$) angular
momentum, for various values of the nuclear charge number~$Z$; the new
calculations that we have performed required the use of massively
parallel computers, and thousands of hours of computing time.
\begin{newA}
  (These numerical data, which have been used for the plots in Figs.\ 
  \ref{fig:5g7}--\ref{fig:deltaGSE}, will be presented in detail
  elsewhere~\cite{lebigot2002b}.)%
\end{newA}
We have also collected the most recent
available values of the self energy.  This provided us with
independent values of the~$A_{60}$ coefficients, extracted from the
numerical self-energies, thus allowing us to check the analytic
results presented in Tables~\ref{tbl:a60P}--\ref{tbl:a60G} (see
Sec.~\ref{sec:check}).

Severe cancellations appeared, between different contributions to
$A_{60}$ (in addition to the cancellation of the $\epsilon$-parameter
divergences): for some of the atomic states investigated, the
absolute magnitude of the $A_{60}$ coefficients is as small as
$10^{-3}$, whereas the largest individual contribution to $A_{60}$,
when following the classification of the corrections according to
Refs.~\cite{jentschura96,JeSoMo1997}, is of the order of $10^{-2}$ or
larger for all atomic states discussed here (see also
Tables~\ref{cancel1} and~\ref{cancel2}).

Future calculations of the Bethe logarithm $\ln
k_0(nl) $ and of the relativistic Bethe
logarithm~$A_{60}(nl_j)$ could also fruitfully be compared to the
estimates given by Eqs.~\refPar{eq:defApproxA}, \refPar{eq:asymptA60}
and \refPar{eq:blCoefs}, and Table~\ref{tbl:asymptCoefs}.
The results presented in this paper also allow one to perform
checks of future exact self-energies obtained by numerical methods, by
comparing their values to the three-term self-energy
approximation~\refPar{approxOrder3} provided here for P and higher-$l$
states.
The values of~$A_{60}$ in Tables~\ref{tbl:a60P}--\ref{tbl:a60G} can be of
interest for analyzing the Lamb shift of highly-excited (high-$n$ and
high-$l$) electronic states in
recent~\cite{BeEtAl1997,schwob99,schwob99b,devries2002} and future
high-precision spectroscopy experiments.
The results of Sect.~\ref{ResHighLow}--\ref{sec:approxA60andBL} also
provide the best available self-energy approximation for many states
$nl_j$ and nuclear charge numbers~$Z$ (see Sec.~\ref{sec:check});
these approximations can for instance be useful in evaluating the
contribution of QED effects in
atoms~\cite{feldman90,parente94,beck99,sugar89} or
molecules~\cite{pyykko2001}.

\typeout{Acknowledgments}
\begin{acknowledgments}
  The authors would like to acknowledge helpful discussions with K.
  Pachucki and J. Sims.  We also thank the CINES (Montpellier, France)
  and the IDRIS (Orsay, France) for grants of time on parallel
  computers (IBM SP2 and SP3~\cite{Disclaimer}).  E.O.L. acknowledges
  support from a Lavoisier fellowship of the French Ministry of
  Foreign Affairs, and support by NIST\@.  U.D.J.  acknowledges
  support from the Deutscher Akademischer Austauschdienst (DAAD)\@.
  G.S.~acknowledges support from BMBF, DFG and from GSI\@.  The
  Kastler Brossel laboratory is Unit\'e Mixte de Recherche~8552 of the
  CNRS\@.
\end{acknowledgments}

\appendix

\typeout{Appendix: Local fits}

\section*{Appendix: local fits}

This appendix describes a fitting procedure which is designed to
extract ``\emph{local}'' numerical quantities from a set of data
points, and to allow one to assess the numerical uncertainty
associated to these quantities. A partial sketch of this procedure was
first introduced in Ref.~\cite{mohr75}.  Here, ``local'' refer for
instance to the evaluation of a perturbation expansion about one
abscissa; the purpose of the method presented here is to perform fits
that are local to an abscissa of interest, as opposed to finding the
best global fit of some data points.  We thus used it in order to
obtain asymptotic coefficients for~$A_{60}(nl_j)$ for P and D states
in Sec.~\ref{sec:a60higherN} (see Table~\ref{tbl:asymptCoefs}), as
well as the asymptotic expansion of the Bethe logarithm $\ln k_0(nl)$
in Eq.~\refPar{eq:blCoefs}---in these applications, the quantities
evaluated are local to either $n=\infty$ or $l=\infty$.  This method
can in principle be applied to many other problems that require local
fits.

In order to describe the local-fit procedure, we take the evaluation
of the limit 
\begin{equation}\label{eq:limitExample}
\lim_{l\rightarrow\infty} l^3\times
\ln k_0(\bar{n}l) 
\end{equation}
as an example---here, we have
$\bar{n}\defi l+1$ and $\ln k_0({n}l) $ is the Bethe
logarithm~\refPar{bethelog}. This limit was
evaluated as $-0.056853(2)$ [see Fig.~\ref{fig:fitBetheLog} and
Eq.~\refPar{eq:blCoefs}].

\begin{figure}
\includegraphics[width=0.7\linewidth]{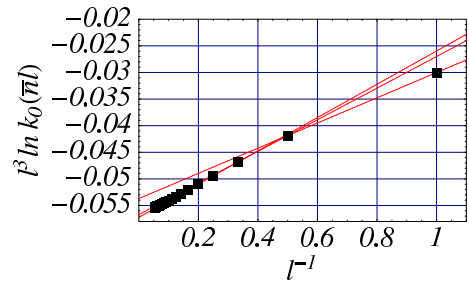}
\caption{\label{fig:fit2ndPass}This figure shows the lines going through a few pairs of successive data points~\refPar{eq:dataSet1}---see also Fig.~\ref{fig:fitBetheLog}. Each of these lines is a local approximation to the curve underlying the data points.  Each line yields an estimate of the limit~\refPar{eq:limitExample} of the data points as $l^{-1}\rightarrow 0$ (this estimate is at the intersection of the line with the $l^{-1}=0$ axis). Fig.~\ref{fig:fit2ndPassLimits} graphically displays these estimates.}
\end{figure}

Figures~\ref{fig:fitBetheLog} and~\ref{fig:fit2ndPass} contain data
points which are relevant to~\refPar{eq:limitExample}: we have plotted
\begin{equation}\label{eq:dataSet1}
l^3\times
\ln k_0(\bar{n}l) 
\end{equation}
as a function of~$l^{-1}$ (with values of the Bethe logarithm found in
Ref.~\cite{drake90note}). The limit~\refPar{eq:limitExample} can visually
be estimated from the data points in Fig.~\ref{fig:fitBetheLog} to
be~$-0.057(1)$.

\begin{figure}
\includegraphics[width=0.7\linewidth]{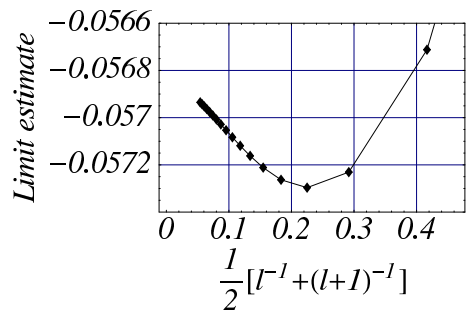}
\caption{\label{fig:fit2ndPassLimits}This figure shows the estimates of limit~\refPar{eq:limitExample} obtained through the two-point fits of Fig.~\ref{fig:fit2ndPass}. From this graph, we  limit~\refPar{eq:limitExample} to be~$-0.0568(1)$, which is more precise than, and coherent with the value~$-0.057(1)$ obtained from the original data points~\refPar{eq:dataSet1} in Figs.~\ref{fig:fitBetheLog} and~\ref{fig:fit2ndPass}.
  The limit estimates are plotted along the vertical direction, while
  the abscissa associated to an estimate is the average abscissa of
  the two data points of Fig.~\ref{fig:fit2ndPass} that were used in
  producing it.}
\end{figure}

In order to improve over the estimate~$-0.057(1)$
for~\refPar{eq:limitExample}, we fit (exactly) each pair of two
consecutive points~\refPar{eq:dataSet1} in Fig.~\ref{fig:fitBetheLog}
with a line%
, as
depicted in Fig.~\ref{fig:fit2ndPass}. Each of the fitting lines in
Fig.~\ref{fig:fit2ndPass} gives an estimate of
limit~\refPar{eq:limitExample} by extrapolation to~$l^{-1}=0$
(intersection of the line with the $l^{-1}=0$~axis).
Figure~\ref{fig:fit2ndPassLimits} contains each of these estimates, as
a function of the average abscissa of the two points that were used in
obtaining it. Because the curve in Fig.~\ref{fig:fit2ndPassLimits} is
relatively flatter than the curve in Fig.~\ref{fig:fit2ndPass}, we can
estimate limit~\refPar{eq:limitExample} with an improved uncertainty;
thus, we deduce from Fig.~\ref{fig:fit2ndPassLimits} the
value~$-0.0568(1)$ for the limit~\refPar{eq:limitExample} that we are
studying, which is coherent with the previous estimate~$-0.057(1)$.

\begin{figure}
\includegraphics[width=0.8\linewidth]{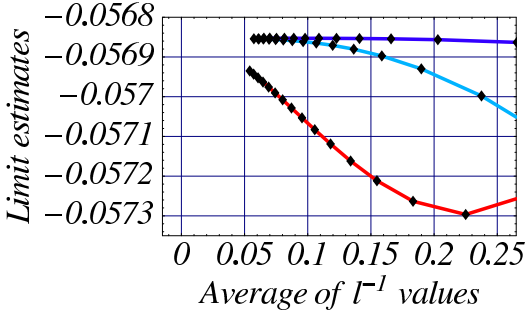}
\caption{\label{fig:multipleFits}
  From the lower to the higher curve: estimates of
  limit~\refPar{eq:limitExample} obtained through fits of the data
  points~\refPar{eq:dataSet1} with polynomials of degree~$1$ (see also
  Fig.~\ref{fig:fit2ndPassLimits}), $3$ and~$5$ (see also
  Fig.~\ref{fig:fit6points}).  Fitting the data
  points~\refPar{eq:dataSet1} of Fig.~\refPar{fig:fit2ndPass} with 1
  to~6 points yielded mutually coherent estimates of
  limit~\refPar{eq:limitExample} with an exponentially decreasing
  error.}
\end{figure}

\begin{figure}
\includegraphics[width=0.8\linewidth]{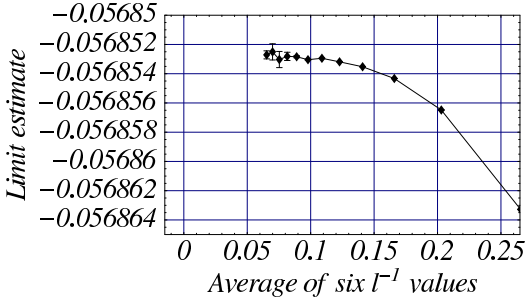}
\caption{\label{fig:fit6points}This figure shows estimates of limit~\refPar{eq:limitExample} obtained by fitting the data points~\refPar{eq:dataSet1} in Fig.~\ref{fig:fit2ndPass} with fifth-degree polynomials (in $l^{-1}$). The high relative stability of the estimates as $l^{-1}\rightarrow 0$ allowed us to give the precise value
  $-0.056853(2)$ in Eq.~\refPar{eq:blCoefs} for
  limit~\refPar{eq:limitExample}.}
\end{figure}

\begin{figure}
\includegraphics[width=0.8\linewidth]{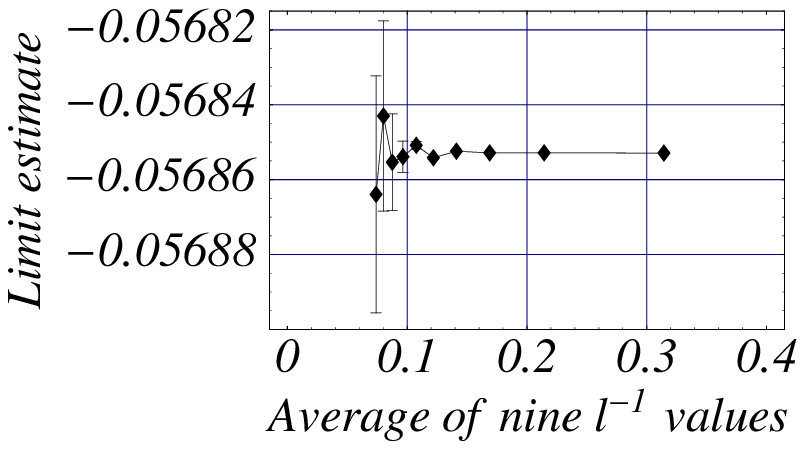}
\caption{\label{fig:fit9points}This figure displays estimates of limit~\refPar{eq:limitExample} obtained by fitting the data points~\refPar{eq:dataSet1} in Fig.~\ref{fig:fit2ndPass} with a eigth-degree polynomials (in $l^{-1}$). It should be compared to Fig.~\ref{fig:fit6points}, which gives a more accurate estimate of limit~\refPar{eq:limitExample} by fitting sequences of only six data points. The accuracy of the local fits performed here first increases with the order of the local approximations to the data points~\refPar{eq:dataSet1} (see Fig.~\ref{fig:multipleFits}), and then eventually decreases (compare this plot to Fig.~\ref{fig:fit6points}).}
\end{figure}

This better estimate~$-0.0568(1)$ of limit~\refPar{eq:limitExample}
can be further improved by continuing to increase the number~$p$ of data
points~\refPar{eq:dataSet1} included in local fits of the data. Thus,
for an increasing number~$p$ of data points, we fitted (exactly) each
set of~$p$ successive points~\refPar{eq:dataSet1} in
Fig.~\ref{fig:fit2ndPass} with a polynomial of degree~$p-1$ (linear
combination of the functions $1$, $l^{-1}$,\ldots, $l^{-(p-1)}$), and
represented the value of the polynomial extrapolated to~$l^{-1} = 0$ as a
function of the average abscissa of the $p$~points.
Fig.~\ref{fig:multipleFits} depicts this process.  The plotted values
are estimates of the limit~\refPar{eq:limitExample} obtained with
higher and higher-order (local) fits of the data
points~\refPar{eq:dataSet1}. In Fig.~\ref{fig:multipleFits}, the abscissa of each estimate is the average of the abscissas~$l^{-1}$ of the fitted data points~\refPar{eq:dataSet1}. We observed that the curves so obtained
become \emph{exponentially flat}, in the sense that their relative
amplitude become exponentially smaller and smaller---until the
uncertainties of individual estimates become important, as described
below.  This fact, which is illustrated in
Fig.~\refPar{fig:multipleFits}, allowed us to obtain more and more
accurate estimates of limit~\refPar{eq:limitExample}.

The most accurate value that we obtained for
limit~\refPar{eq:limitExample} through the local-fit procedure
described here is $-0.056853(2)$ [see Eq.~\refPar{eq:blCoefs}], as is
illustrated in Fig.~\ref{fig:fit6points}. This limit was obtained by
fitting each sequence of $p=6$ data points with a fifth-degree
polynomial. Fits of the data points~\refPar{eq:dataSet1} with larger
numbers of data points display more irregular estimate curves; this
can for instance be seen by comparing Fig.~\ref{fig:fit6points} with
Fig.~\ref{fig:fit9points}.

As we have seen above, the uncertainty in the fitted value can be
evaluated by visually prolongating the fitting curves (i.e., curves
such as those of
Figs.~\ref{fig:fit2ndPassLimits}--\ref{fig:fit9points}). Another
uncertainty must in general be taken into account in order to obtain a
reliable estimate for the fitted quantity: the uncertainty in the data
points.  All the curves presented in this appendix do contain error
bars that reflect the uncertainties in the estimates
of~\refPar{eq:limitExample} that come from the uncertainties in the
data points~\refPar{eq:dataSet1}.  We evaluated the uncertainty
associated to each fit of $p$~data points~\refPar{eq:dataSet1} by
calculating three fits: a fit with the middle values of the ordinates, a fit
with the higher values, and a fit with the lower values; the three
estimates of the fitted quantity~\refPar{eq:limitExample} obtained
through this procedure define an estimate with an error bar (see,
e.g., Fig.~\ref{fig:fit9points}). Other ways of estimating the
uncertainty in the fit result can be used; a good choice of
uncertainty evaluation yields successive estimates of the fitted
quantity that are compatible with a smooth curve of estimates [see,
e.g., Fig.~\ref{fig:fit9points}, where the less precise estimates of
limit~\refPar{eq:limitExample} lie in the prolongation of the more
precise values, which are on the right of the plot].

One of the advantages of the local-fit method presented in this
appendix is that data points that are located far from the abscissa of
interest ($l^{-1}=0$, here) can fruitfully be used in evaluating the
fitted quantity [limit~\refPar{eq:limitExample}, in our example].
Thus, as Fig.~\ref{fig:fit9points} illustrates, data
points~\refPar{eq:dataSet1} with ``large'' abscissas can yield more
precise estimates of limit~\refPar{eq:limitExample} than data points
with small abscissas. This behavior is particularly useful when data
points in the region of interest have relatively large uncertainties.

The procedure detailed in this Appendix also allows one to study the
quality of lists of numerical results that should lie on a smooth
curve, but whose coherence is not obvious through a simple inspection
or plot of the values. In fact, curves such as those found in
Figs.~\ref{fig:fit2ndPassLimits}--\ref{fig:fit9points} can be very
sensitive to small errors in a list of numerical values. We have not
noticed such errors in the $A_{60}$ values of Tables~\ref{tbl:a60P}
and~\ref{tbl:a60D} while evaluating the asymptotic coefficients
reported in Table~\ref{tbl:asymptCoefs}; this provided an additional
check of the values reported in these tables (see also Sec.~\ref{sec:check}).

The local-fit method described here is not restricted to the
asymptotic study of the Bethe logarithm that we have used as an
example.  In general, it can yield precise estimates of quantities
that are local to a set of data point [such as
limit~\refPar{eq:limitExample}], including, for instance, perturbation
coefficients of non-analytic expansions [e.g., Eq.~\refPar{defFLO}].

\newcommand{\noopsort}[1]{}


\begin{thebibliography}{65}
\bibitem[{Dis()}]{Disclaimer}
\bibinfo{howpublished}{Certain commercial equipment, instruments, or materials
  are identified in this paper to foster understanding. Such identification
  does not imply recommendation or endorsement by the National Institute of
  Standards and Technology, nor does it imply that the materials or equipment
  identified are necessarily the best available for the purpose.}
\end{thebibliography}
\end{document}